\documentclass[12pt,paper]{article}
\usepackage{avant}
\usepackage[T1]{fontenc}
\usepackage[latin9]{inputenc}
\usepackage{array}
\usepackage{float}
\usepackage{verbatim}
\usepackage{multirow}
\usepackage{amsmath}
\usepackage{amssymb}
\usepackage{graphicx}
\usepackage{acronym}
\usepackage{multirow}
\usepackage{setspace}
\usepackage{vmargin}
\setmarginsrb{1.0in}{0.5in}{1.0in}{1.5in}{0.5in}{0.5in}{0.5in}{0.5in}
\usepackage{hyperref}
\makeatletter

\providecommand{\tabularnewline}{\\}
\floatstyle{ruled}
\newfloat{algorithm}{tbp}{loa}
\providecommand{\algorithmname}{Algorithm}
\floatname{algorithm}{\protect\algorithmname}

\usepackage{avant}
\usepackage[T1]{fontenc}
\usepackage[latin9]{inputenc}
\usepackage{amsmath}
\usepackage{amssymb}

\makeatletter

\providecommand{\tabularnewline}{\\}

\@ifundefined{definecolor}{\usepackage{color}}{}
\RequirePackage[displaymath]{lineno}

\@ifundefined{definecolor}{\usepackage{color}}{}
\usepackage{amsfonts}\usepackage{latexsym}

\usepackage{mathrsfs}\usepackage{amsthm}\usepackage{latexsym}\usepackage{booktabs}\@ifundefined{definecolor}{\usepackage{color}}{}

\usepackage{times}\usepackage{avant}\@ifundefined{definecolor}{\usepackage{color}}{}

\numberwithin{equation}{section}
\numberwithin{figure}{section}

\newcommand{\indep}{\;\, \rule[0em]{.03em}{.67em} \hspace{-.25em}
\rule[0em]{.65em}{.03em} \hspace{-.25em}
\rule[0em]{.03em}{.67em}\;\,}

\newtheorem{pn}{Proposition}

\newtheorem{defi}{Definition}

\usepackage[english]{babel}

\makeatother

\begin{document}
{\setlength{\baselineskip}{1.5\baselineskip} \global\long\def\mbx{\mathbf{x}}
\global\long\def\J{D}
\global\long\def\d{d}
 \global\long\def\oneD{1D}
 \global\long\def\mbZ{\mathbf{Z}}
 \global\long\def\mbX{\mathbf{X}}
 \global\long\def\mbz{\mathbf{z}}
 \global\long\def\mbA{\mathbf{A}}
 \global\long\def\mbell{\mathbf{\ell}}
 \global\long\def\mbc{\mathbf{c}}
 \global\long\def\mbD{\mathbf{D}}
 \global\long\def\mbB{\mathbf{B}}
 \global\long\def\mbC{\mathbf{C}}
 \global\long\def\mbF{\mathbf{F}}
 \global\long\def\mbzero{\mathbf{0}}
 \global\long\def\mbI{\mathbf{I}}
 \global\long\def\mbS{\mathbf{S}}
 \global\long\def\mbR{\mathbf{R}}
 \global\long\def\mbU{\mathbf{U}}
 \global\long\def\mbV{\mathbf{V}}
 \global\long\def\mbW{\mathbf{W}}
 \global\long\def\mbY{\mathbf{Y}}
 \global\long\def\mby{\mathbf{y}}
 \global\long\def\mbP{\mathbf{P}}
 \global\long\def\mbQ{\mathbf{Q}}
 \global\long\def\mbO{\mathbf{O}}
 \global\long\def\mbT{\mathbf{T}}
 \global\long\def\mbM{\mathbf{M}}
 \global\long\def\hatmbM{\widehat{\mathbf{M}}}
 \global\long\def\hatmbA{\widehat{\mbA}}
 \global\long\def\hatmbB{\widehat{\mbB}}
 \global\long\def\hatmbU{\widehat{\mathbf{U}}}
 \global\long\def\hatmbB{\widehat{\mathbf{B}}}
 \global\long\def\mbK{\mathbf{K}}
 \global\long\def\bolh{\boldsymbol{h}}
 \global\long\def\mbE{\mathbf{E}}
 \global\long\def\mbJ{\mathbf{J}}
 \global\long\def\mbu{\mathbf{u}}
 \global\long\def\mbw{\mathbf{w}}
 \global\long\def\mbd{\mathbf{d}}
 \global\long\def\mbs{\mathbf{s}}
 \global\long\def\mbv{\mathbf{v}}
 \global\long\def\mbt{\mathbf{t}}
 \global\long\def\mbr{\mathbf{r}}
 \global\long\def\bollambda{\boldsymbol{\lambda}}
 \global\long\def\mbh{\mathbf{h}}
 \global\long\def\hatmbh{\widehat{\mbh}}
 \global\long\def\mbho{\mathbf{h}_{0}}
 \global\long\def\bolGamma{\boldsymbol{\Gamma}}
 \global\long\def\bolGammao{\bolGamma_{0}}
 \global\long\def\mbN{\mathbf{N}}
 \global\long\def\hatbolh{\widehat{\bolh}}
 \global\long\def\hatvartheta{\widehat{\vartheta}}

\global\long\def\mbbR{\mathbb{R}}
 \global\long\def\mbbS{\mathbb{S}}
 \global\long\def\mbbX{\mathbb{X}}
 \global\long\def\mbbY{\mathbb{Y}}
 \global\long\def\mba{\mathbf{a}}
 \global\long\def\mbb{\mathbf{b}}
 \global\long\def\mbe{\mathbf{e}}
 \global\long\def\mbf{\mathbf{f}}
 \global\long\def\calE{\mathcal{E}}
 \global\long\def\bolTheta{\boldsymbol{\Theta}}
 \global\long\def\bolSigmares{\boldsymbol{\Sigma}_{\mathrm{res}}}
 \global\long\def\bolSigmafit{\boldsymbol{\Sigma}_{\mathrm{fit}}}
 \global\long\def\boltheta{\boldsymbol{\theta}}
 \global\long\def\sqW{\sqrt{W}}

\global\long\def\bolSigma{\boldsymbol{\Sigma}}
 \global\long\def\hatbolSigma{\widehat{\bolSigma}}
 \global\long\def\bolxi{\boldsymbol{\xi}}
 \global\long\def\hatbolxi{\widehat{\boldsymbol{\xi}}}
 \global\long\def\bolphi{\boldsymbol{\phi}}
 \global\long\def\hatboltheta{\widehat{\boltheta}}
 \global\long\def\hatbolphi{\widehat{\boldsymbol{\phi}}}
 \global\long\def\bolpsi{\boldsymbol{\psi}}
 \global\long\def\hatbolpsi{\widehat{\boldsymbol{\psi}}}
 \global\long\def\boleta{\boldsymbol{\eta}}
 \global\long\def\tilboleta{\tilde{\boleta}}
 \global\long\def\hatboleta{\widehat{\boleta}}
 \global\long\def\bolbeta{\boldsymbol{\beta}}
 \global\long\def\hatbolbeta{\widehat{\bolbeta}}
 \global\long\def\hatbolbetaenv{\widehat{\bolbeta}_{\mathrm{env}}}
 \global\long\def\bolGamma{\boldsymbol{\Gamma}}
 \global\long\def\hatbolGamma{\widehat{\bolGamma}}
 \global\long\def\bolPhi{\boldsymbol{\Phi}}
 \global\long\def\tilbolPhi{\tilde{\bolPhi}}
 \global\long\def\bolPhio{\boldsymbol{\Phi}_{0}}
 \global\long\def\bolOmega{\boldsymbol{\Omega}}
 \global\long\def\bolOmegao{\boldsymbol{\Omega}_{0}}
 \global\long\def\invbolGamma{\bolGamma^{-1}}
 \global\long\def\invhatbolGamma{\hatbolGamma^{-1}}
 \global\long\def\bolnu{\boldsymbol{\nu}}
 \global\long\def\hatbolnu{\widehat{\bolnu}}
 \global\long\def\bolgamma{\boldsymbol{\gamma}}
 \global\long\def\hatbolgamma{\widehat{\bolgamma}}
 \global\long\def\bolzeta{\boldsymbol{\zeta}}
 \global\long\def\bolvarepsilon{\boldsymbol{\varepsilon}}
 \global\long\def\hatbolzeta{\widehat{\bolzeta}}
 \global\long\def\bolmu{\boldsymbol{\mu}}
 \global\long\def\hatbolmu{\widehat{\bolmu}}
 \global\long\def\bolepsilon{\boldsymbol{\epsilon}}
 \global\long\def\bolSigmaX{\bolSigma_{\mbX}}
 \global\long\def\bolSigmaY{\bolSigma_{\mbY}}
 \global\long\def\bolSigmaXY{\boldsymbol{\Sigma}_{\mbX\mbY}}
 \global\long\def\mbSX{\mathbf{S}_{\mbX}}
 \global\long\def\mbSY{\mathbf{S}_{\mbY}}
 \global\long\def\mbSXY{\mathbf{S}_{\mbX\mbY}}
 \global\long\def\mbSYX{\mathbf{S}_{\mbY\mbX}}
 \global\long\def\mbRYX{\mathbf{S}_{\mbY|\mbX}}
 \global\long\def\mbRXY{\mathbf{S}_{\mbX|\mbY}}
 \global\long\def\bolalpha{\boldsymbol{\alpha}}
 \global\long\def\mbSc{\mbS_{\mbC}}
 \global\long\def\mbSd{\mbS_{\mbD}}
 \global\long\def\bolrho{\boldsymbol{\rho}}
 \global\long\def\bolPsi{\boldsymbol{\Psi}}
 \global\long\def\calT{\mathcal{T}}
 \global\long\def\calW{\mathcal{W}}
 \global\long\def\calD{\mathcal{D}}
 \global\long\def\calU{\mathcal{U}}
 \global\long\def\calC{\mathcal{C}}
 \global\long\def\bolsigma{\boldsymbol{\sigma}}

\global\long\def\sumn{\sum_{i=1}^{n}}
 \global\long\def\sumnj{\sum_{j=1}^{n}}
 \global\long\def\suminf{\sum_{k=1}^{\infty}}
 \global\long\def\bolSigmaYonX{\bolSigma_{\mbY|\mbX}}
 \global\long\def\bolSigmac{\bolSigma_{\mbC}}
 \global\long\def\bolSigmad{\bolSigma_{\mbD}}
 \global\long\def\bolmux{\bolmu_{\mbX}}
 \global\long\def\bolmuy{\bolmu_{\mbY}}
 \global\long\def\bolmuc{\bolmu_{\mbc}}
 \global\long\def\bolDelta{\boldsymbol{\Delta}}
 \global\long\def\bolvarphi{\boldsymbol{\varphi}}
 \global\long\def\boldelta{\boldsymbol{\delta}}
 \global\long\def\tilbolDelta{\tilde{\bolDelta}}
 \global\long\def\tilboldelta{\tilde{\boldelta}}
 \global\long\def\bolLambda{\boldsymbol{\Lambda}}

\global\long\def\vecc{\mathrm{vec}}
 \global\long\def\Prob{\mathrm{Pr}}
 \global\long\def\E{\mathrm{E}}
 \global\long\def\Cov{\mathrm{cov}}
 \global\long\def\Corr{\mathrm{corr}}
 \global\long\def\F{\mathrm{F}}
 \global\long\def\Var{\mathrm{var}}
 \global\long\def\dimension{\mathrm{dim}}
 \global\long\def\spn{\mathrm{span}}
 \global\long\def\vech{\mathrm{vech}}

\global\long\def\Syx{\mathcal{S}_{Y|\mbX}}
 \global\long\def\bolGammapsi{\bolGamma_{\hatbolpsi}}
 \global\long\def\hatbolGammapsi{\widehat{\bolGamma}_{\hatbolpsi}}
 \global\long\def\dx{d_{X}}
 \global\long\def\dy{d_{Y}}
 \global\long\def\fp{f_{p}}
 \global\long\def\fx{f_{x}}

\global\long\def\calH{{\cal H}}
 \global\long\def\calB{{\cal B}}
 \global\long\def\calV{{\cal V}}
 \global\long\def\calS{{\cal S}}
 \global\long\def\calR{{\cal R}}
 \global\long\def\calRp{\calR^{\perp}}
 \global\long\def\calSp{\calS^{\perp}}
 \global\long\def\calL{\mathcal{L}}
 \global\long\def\calA{\mathcal{A}}
 \global\long\def\boltheta{\boldsymbol{\theta}}
 \global\long\def\bolTheta{\boldsymbol{\Theta}}
 \global\long\def\calr{\mathcal{r}}

\global\long\def\KenvS{\mathcal{E}_{K}(\calS)}
 \global\long\def\KenvB{\mathcal{E}_{K}(\mathcal{B})}
 \global\long\def\Xenv{\mathcal{E}_{\bolSigma_{\mbX}}(\mathcal{L})}
 \global\long\def\Yenv{\mathcal{E}_{\bolSigma_{\mbY}}(\mathcal{R})}
 \global\long\def\YonXenv{\mathcal{E}_{\bolSigmaYonX}(\mathcal{R})}
 \global\long\def\MenvB{\calE_{\mbM}(\calB)}
 \global\long\def\calM{\mathcal{M}}
 \global\long\def\calG{\mathcal{G}}
 \global\long\def\Ienv{\calE_{\mbV}}
 \global\long\def\WLS{\mathrm{WLS}}
 \global\long\def\hatbolalpha{\widehat{\bolalpha}}
 \global\long\def\Env{\mathrm{env}}

\global\long\def\tr{\mathrm{trace}}
 \global\long\def\dg{\mathrm{diag}}
 \global\long\def\asyVar{\mathrm{avar}}
 \global\long\def\mbbC{\mathbb{C}}

\global\long\def\mbG{\mathbf{G}}
 \global\long\def\mbH{\mathbf{H}}

\global\long\def\mbGo{\mathbf{G}_{0}}
 \global\long\def\mbHo{\mathbf{H}_{0}}
 \global\long\def\mbg{\mathbf{\mathbf{g}}}
 \global\long\def\tildembG{\widetilde{\mbG}}

\global\long\def\mbL{\mathbf{L}}
 \global\long\def\mbR{\mathbf{R}}
 \global\long\def\mbLo{\mathbf{L}_{0}}
 \global\long\def\mbRo{\mathbf{R}_{0}}
 \global\long\def\bolell{\boldsymbol{\ell}}
 \global\long\def\RRR{\mathrm{RRR}}
 \global\long\def\ERRR{\mathrm{ERR}}
 \global\long\def\OLS{\mathrm{OLS}}
  \global\long\def\FG{FG}

\global\long\def\tilbolGamma{\widetilde{\bolGamma}}
 \global\long\def\tilbolGammao{\widetilde{\bolGamma}_{0}}
 \global\long\def\hatbolOmega{\widehat{\bolOmega}}
 \global\long\def\hatnu{\widehat{\nu}}
 \global\long\def\hatomega{\widehat{\omega}}


\acrodef{GPLS}[$\mathrm{GPLS}$]{}
\acrodef{oneD}[1D]{}


\title{{Algorithms for Envelope Estimation}}

\author{R. Dennis Cook%
\thanks{{\small R. Dennis Cook is Professor, School of Statistics, University
of Minnesota, Minneapolis, MN 55455 (E-mail: dennis@stat.umn.edu).}%
} \ and\ Xin Zhang%
\thanks{{\small Xin Zhang is Ph.D student, School of Statistics, University
of Minnesota, Minneapolis, MN, 55455 (Email: zhan0648@umn.edu). }%
} }
\maketitle
\begin{abstract}
Envelopes were recently proposed as methods for reducing estimative variation in multivariate linear regression.
Estimation of an envelope usually involves optimization over Grassmann manifolds.
We propose a fast and widely applicable one-dimensional (\oneD)\ algorithm for estimating an envelope in general.
We reveal an important structural property of envelopes that facilitates our algorithm, and we prove both Fisher consistency and $\sqrt{n}$-consistency of the algorithm. \\

\end{abstract}

\section{Introduction}
Envelope methods aim to reduce estimative variation in multivariate linear models.
The reduction is typically associated with predictors or responses, and can generally be interpreted as effective dimensionality reduction in the parameter space.
Such reduction is achieved by enveloping the variation in the data that is material to the goals of the analysis while simultaneously excluding the immaterial variation.
Efficiency gains are then achieved by essentially basing estimation on the material variation alone.
The improvement in estimation and prediction can be quite substantial when the immaterial variation is large,  sometimes equivalent to taking thousands of additional observations.

The novel notion of an envelope, which is a subspace of the predictor or response spaces containing all of the material variation, was first introduced by Cook et al. (2010) for response reduction in multivariate linear models, subsequently
studied by Su and Cook (2011) for partial reduction and recently studied
by Cook et al. (2013) for predictor reduction.
In particular, Cook et al. (2013)
found that the commonly used PLS algorithm, SIMPLS (de Jong 1993), is in fact based on a $\sqrt{n}$-consistent envelope estimator, while the corresponding likelihood-based approach produces a better estimator.

The likelihood-based approach to envelope estimation requires, for a given envelope dimension $u$, optimizing an objective function of the form $f(\bolGamma)$, where $\bolGamma$ is a $k\times u$, $k>u$, semi-orthogonal basis matrix for the envelope.
The objective function satisfies $f(\bolGamma) = f(\bolGamma\mbO)$ for any $u\times u$ orthogonal matrix $\mbO$. Hence the optimization is essentially over the set of all $u$-dimensional subspaces of $\mbbR^{k}$, which is a Grassmann manifold denoted as $\calG_{u,k}$.
Since $u(k-u)$ real numbers are  required to specify an element of  $\calG_{u,k}$ uniquely, the optimization is essentially over $u(k-u)$ real dimensions.  In multivariate linear regression, $k$ can be either the number of responses $r$ or the number of predictors $p$, depending on  whether one is pursuing response or predictor reduction.

All present envelope methods rely on the Matlab package  {\tt sg\_min}  by Ross A. Lippert (\url{http://web.mit.edu/~ripper/www/software/}) to optimize $f(\bolGamma)$. This package provides iterative optimization techniques on Stiefel and Grassmann manifolds, including non-linear conjugate gradient (PRCG and FRCG) iterations, dog-leg steps and Newton's method. To implement an envelope estimation procedure, one needs to specify the objective function $f(\bolGamma)$ and its analytical first-order derivative function. Then given an initial value of $\bolGamma$, this package will compute numerical second-order derivatives and iterate until convergence or the maximum number of iterations is reached. The Matlab toolbox {\tt envlp} by R. D. Cook, Z. Su and Y. Yang (\url{http://code.google.com/p/envlp/}) uses {\tt sg\_min} to implement a variety of envelope estimators along with associated inference methods.  The {\tt sg\_min} package works well for envelope estimation, but nevertheless, optimization is often computationally difficult for large values of $u(k-u)$.
At higher dimensions, each iteration becomes exponentially slower, local minima can become a serious issue and good starting values are essential.  The {\tt envlp} toolbox implements a seemingly different version of $f(\bolGamma)$ for each type of envelope, along with tailored starting values.

In this article we present two advances in envelope computation.  First,  we propose in Section~\ref{sec:objectiveFun} a model-free objective function $J_n(\bolGamma)$ for estimating an envelope and
show that the three major envelope methods are based on special cases of $J_{n}$.  This unifying objective function is to be optimized over the Grassmann manifold $\calG_{u,k}$, which for larger values of $u(k-u)$ will be subject to the same computational limitations associated with speed, local minima and starting values.
Second, we propose in Section~\ref{sec:1Dalgorithm}  a fast one-dimensional (1D) algorithm that mitigates these computational issues.
To adapt the envelope construction for relatively large values of $u(k-u)$, we break down Grassmann optimization into a series of one-dimensional optimizations so that the estimation procedure is speeded up greatly, and starting values and local minima are no longer an issue.
Although it may be impossible to break down a general $u$-dimensional Grassmann optimization problem, we rely on special characteristics of envelopes in statistical problems to achieve the breakdown of envelope estimation.   The resulting \oneD\ algorithm, which is easy-to-implement, stable and requires no initial value input, can be tens to hundreds times faster than the general Grassmann manifold optimization for $u>1$, while still providing a desirable $\sqrt{n}$-consistent envelope estimator.
Very recently, Cook and Zhang (2014) introduced simultaneous reduction of the predictors and the response by envelopes.
The objective function in Cook and Zhang (2014) has the form of $f(\mbL,\mbR)$ where $\mbL$ and $\mbR$ are both semi-orthogonal matrices and the optimization is over two Grassmann manifolds.
They used special forms of the \oneD\ algorithm to find initial values for $\mbL$ and $\mbR$.
The \oneD\ algorithm we introduce in Section~\ref{sec:1Dalgorithm} is much more general and is directly applicable beyond the multivariate linear regression context.

The rest of this article is organized as follows.
In Section~\ref{sec:review env}, we review briefly key algebraic foundations of envelopes, and also review concepts and methodology in the context of an example.
Because envelopes are nascent methodology, the level of detail in this example is somewhat greater than what might be considered traditional.
Section~\ref{sec:data} consists of simulation studies and a data example to further demonstrate the advantages of the \oneD\ algorithm.
Section~\ref{sec:conclusion} is a brief conclusion of this paper.
Proofs and technical details are included in the Appendix.

The following notations and definitions will be used in our exposition.
Let $\mbbR^{m\times n}$ be the set of all real $m\times n$ matrices
and let $\mbbS^{k}$ be the set of all real and symmetric
$k\times k$ matrices.
Suppose $\mbM\in\mbbR^{m\times n}$, then $\spn(\mbM)\subseteq\mbbR^{m}$
is the subspace spanned by columns of $\mbM$.
We use $\mbP_{\mbA(\mbV)}=\mbA(\mbA^{T}\mbV\mbA)^{-1}\mbA^{T}\mbV$
to denote the projection onto $\spn(\mbA)$ with the $\mbV$ inner
product and use $\mbP_{\mbA}$ to denote projection onto $\spn(\mbA)$
with the identity inner product. Let $\mbQ_{\mbA(\mbV)}=\mbI-\mbP_{\mbA(\mbV)}$.
Sample covariance matrices are represented as $\mbS_{(\cdot)}$ and defined with the divisor $n$.
For instance, $\mbSX=\sumn(\mbX_{i}-\bar{\mbX})(\mbX_{i}-\bar{\mbX})^{T}/n$,  $\mbS_{\mbX\mbY}=\sumn(\mbX_{i}-\bar{\mbX})(\mbY_{i}-\bar{\mbY})^{T}/n$ and $\mbS_{\mbY|\mbX}$ denotes the covariance matrix of the residuals from the linear fit of $\mbY$ on $\mbX$: $\mbS_{\mbY|\mbX}=\mbSY-\mbS_{\mbY\mbX}\mbS^{-1}_{\mbX}\mbS_{\mbX\mbY}$.

\section{Review of envelopes \label{sec:review env}}

\subsection{Definition of an envelope}

This following definition of a reducing subspace is equivalent to the usual
definition found in functional analysis (Conway 1990) and in the
literature on invariant subspaces, but the underlying notion of reduction
is incompatible with how it is usually understood in statistics.
Nevertheless, it is common terminology in those areas and is the basis for the definition of an envelope (Cook, et al., 2010) which is central to our developments.
\begin{defi}\label{def: reducing subspace} A subspace $\calR\subseteq\mbbR^{\d}$
is said to be a reducing subspace of $\mbM\in\mbbR^{\d\times \d}$ if
$\calR$ decomposes $\mbM$ as $\mbM=\mbP_{\calR}\mbM\mbP_{\calR}+\mbQ_{\calR}\mbM\mbQ_{\calR}$.
If $\calR$ is a reducing subspace of $\mbM$, we say that $\calR$
reduces $\mbM$.
\end{defi}

The next definition shows how to construct an envelope in terms of reducing subspaces.

\begin{defi}\label{def: CLC env} Let $\mbM\in\mbbS^{\d}$
and let $\calB\subseteq\spn(\mbM)$. Then the $\mbM$-envelope of
$\calB$, denoted by $\calE_{\mbM}(\calB)$, is the intersection of
all reducing subspaces of $\mbM$ that contain $\calB$.
\end{defi}

The intersection of two reducing subspaces of $\mbM$ is still a reducing
subspace of $\mbM$. This means that $\calE_{\mbM}(\calB)$, which
is unique by its definition, is the smallest reducing subspace containing
$\calB$. Also, the $\mbM$-envelope of $\calB$ always exist because
of the requirement $\calB\subseteq\spn(\mbM)$. If $\spn(\mbU)=\calB$,
then we write $\calE_{\mbM}(\mbU):=\calE_{\mbM}(\spn(\mbU))=\calE_{\mbM}(\calB)$
to avoid notation proliferation. Let $\calE_{\mbM}^{\perp}(\mbU)$
denote the orthogonal complement of $\calE_{\mbM}(\mbU)$.

The following proposition from Cook, et al. (2010) gives a characterization
of envelopes.
\begin{pn}\label{prop:CLC env} Let $q \leq \d$ denote the number of eigenspaces of $\mbM\in\mbbS^{\d}$.
 Then the $\mbM$-envelope of $\mathcal{B}$
can be constructed as $\mathcal{E}_{\mbM}(\mathcal{B})=\sum_{i=1}^{q}\mbP_{i}\mathcal{B}$,
where $\mbP_{i}$ is the projection onto the $i$-th eigenspace of
$\mbM$.
\end{pn}

From this proposition, we see that the $\mbM$-envelope of $\calB$
is the sum of the eigenspaces of $\mbM$ that are not orthogonal to
$\calB$; that is, the eigenspaces of $\mbM$ onto which $\calB$
projects non-trivially. This implies that the envelope is the span
of some subset of the eigenspaces of $\mbM$. In the regression context,
$\calB$ is typically the span of a regression coefficient matrix
or a matrix of cross-covariances, and $\mbM$ is chosen as a covariance
matrix which is usually positive definite. We next illustrate the
potential gain of envelope method using a linear regression example.

\subsection{Concepts and methodology}\label{sec:concepts}
We use Kenward's (1987) data to illustrate the working mechanism of envelopes in multivariate linear regression.
These data came from an experiment to compare two treatments for the control of an intestinal parasite in cattle.
Thirty animals were randomly assigned to each of the two treatments.
Their weights (in kilograms) were recorded at the beginning of the study prior to treatment application and at 10 times during the study corresponding to weeks 2, 4, 6, ..., 18 and 19; that is, at
 two-weeks intervals except the last which was over a one-week interval.
The goal was to find if there is a detectable difference between the two treatments and, if such a difference exists, the time at which it first occurred.
As emphasized by Kenward (1987), although these data have a typical longitudinal structure, the nature of the disease means that growth during the experiment is not amenable to modeling as a smooth function of time, and that fitting growth profiles with a low degree polynomial may hide interesting features of the data because the mean growth curves for the two treatment groups are very close relative to their variation from animal to animal.  Indeed, profile plots of the data suggest no difference between the treatments.  Kenward modeled the data using a multivariate linear model with an ``ante-dependence'' covariance structure.  Here we proceed with an envelope analysis based on a multivariate linear model, following the structure outlined by Cook et al. (2010).

Neglecting the basal measurement for simplicity, let $\mbY_i\in\mbbR^{10}$, $i=1,\dots,60$, be the vector of weight measurements of each animal over time and let $X_i=0$ or $1$  indicate the two treatments.
Our interest lies in the regression coefficient $\bolbeta$ from the multivariate linear regression $\mbY=\bolalpha+\bolbeta X+\bolepsilon$, where it is assumed that $\bolepsilon\sim N(0,\bolSigma)$.  Let $\hatbolbeta_{\OLS}$ denote the ordinary least squares estimator of $\bolbeta$, which  is also the maximum likelihood estimator.  The estimates and their residual bootstrap standard errors are shown in Table~\ref{tab:cow}.  The maximum absolute $t$-value over the elements of $\hatbolbeta_{\OLS}$ is $1.30$, suggesting that the treatments do not have a differential affect on animal weight.  However, with a value of $26.9$ on $10$ degrees of freedom, the likelihood ratio statistics for the hypothesis $\bolbeta =0$ indicates otherwise.  We next turn to an envelope analysis.

Let $\bolGamma \in \mbbR^{10 \times u}$ be a semi-orthogonal basis matrix for $\mathcal{E}_{\bolSigma}(\bolbeta)$, the $\bolSigma$-envelope of $\spn(\bolbeta)$, and let $(\bolGamma, \bolGamma_{0})$ be an orthogonal matrix.  Then $\spn(\bolbeta) \in \mathcal{E}_{\bolSigma}(\bolbeta)$ and we can express $\bolbeta = \bolGamma \boleta$, where $\boleta \in \mbbR^{u \times 1}$ carries the coordinates of $\bolbeta$ relative to the basis $\bolGamma$ and $1 \leq u \leq 10$. The envelope version of the multivariate linear model can now be written as $\mbY=\bolalpha+\bolGamma\boleta X+\bolepsilon$, with $\bolSigma = \bolGamma \bolOmega \bolGamma^{T} + \bolGamma_{0}\bolOmega_{0}\bolGamma_{0}^{T}$, where $\bolOmega \in \mbbR^{u \times u}$ and $\bolOmega_{0} \in \mbbR^{(10-u) \times (10-u)}$ are positive definite matrices.  Under this model, $\bolGamma_{0}^{T}\mbY|X \sim \bolGamma_{0}^{T}\mbY$ and $\bolGamma_{0}^{T}\mbY \indep \bolGamma^{T}\mbY|X$.  Consequently, $\bolGamma_{0}^{T}\mbY$ does not respond to changes in $X$ either marginally or because of an association with $\bolGamma^{T}\mbY$.  For these reasons we regard $\bolGamma_{0}^{T}\mbY$ as the immaterial information and $\bolGamma^{T}\mbY$ as the material information.  Envelope analyses are particularly effective when the immaterial variation $\Var(\bolGamma_{0}^{T}\mbY)$ is large relative to the material variation $\Var(\bolGamma^{T}\mbY)$. After finding a value $\hatbolGamma$ of $\bolGamma$ that minimizes the likelihood-based Grassmann objective function $\log|\bolGamma^T\mbS^{-1}_{\mbY}\bolGamma|
+\log|\bolGamma^T\mbS_{\mbY|X}\bolGamma|$, which will be discussed in Section~\ref{sec:objectiveFun}, over all semi-orthogonal matrices $\bolGamma\in\mbbR^{10\times u}$, the envelope estimator of $\bolbeta$ is given by $\hatbolbeta_{\Env} = \mbP_{\hatbolGamma} \hatbolbeta_{\OLS}$.  Because $u(10-u) \leq 25$, the real dimensions involved in this optimization are small and the {\tt envlp} code can be used without running into computational issues.
Standard methods like BIC and likelihood ratio testing can be used to guide the choice of the envelope dimension $u$. Both methods indicate clearly that $u=1$ in this illustration. In other words, the treatment difference is manifested in only one linear combination $\bolGamma^{T}\mbY$ of the response vector.

The envelope estimate $\hatbolbeta_{\Env}$ is shown in Table~\ref{tab:cow} along with bootstrap standard errors and standard errors obtained from the asymptotic normal distribution of $\sqrt{n}(\hatbolbeta_{\Env} - \bolbeta)$ by the plug-in method (See Cook et al. (2010) for the asymptotic covariance matrix).  We see that the asymptotic standard errors are a bit smaller than the bootstrap standard errors. Using either set of standard errors and using a Bonferroni adjustment for multiple testing,  we see that there is a difference between the treatments and that the difference is first manifested around week 10 and remains thereafter.  As shown in the final row of Table~\ref{tab:cow}, the bootstrap standard errors for the elements of $\hatbolbeta_{\OLS}$  were $2.2$ to $5.9$ times those of $\hatbolbeta_{\Env}$.  Hundreds of additional samples would be needed to reduce the standard errors of the elements of $\hatbolbeta_{\OLS}$ by these amounts.

We  conclude this example by considering the regression of the 6th and 7th element of $\mbY$, corresponding to weeks 12 and 14, on $X$, now letting $\mbY = (Y_{6}, Y_{7})^{T}$.  This allows us to represent the regression graphically and thereby provide intuition on the working mechanism of an envelope analysis.
Figure~\ref{fig:cow} shows a plot of $Y_{6}$ versus $Y_{7}$ with the points marked by treatment.
Since $\bolbeta=\E(\mbY|X=1)-\E(\mbY|X=0) \in \mbbR^{2 \times 1}$, the standard estimator for
$\bolbeta$ is obtained as the difference in the marginal means after projecting the data onto the horizontal and vertical axes of the plot.  The two densities estimates with the larger variation shown along the horizontal axes of the plot represent this operation.  These density estimates are nearly identical, which explains the relatively small $t$-values from the standard model mentioned previously.  However, it is clear from the figure that the treatments do differ.

An envelope analysis infers that $\bolbeta=(\beta_6,\beta_7)^T$ is parallel to the second eigenvector of $\bolSigma=\Cov(Y_{6},Y_7)$.
Hence by Proposition~\ref{prop:CLC env}, $\calE_{\bolSigma}(\bolbeta)=\spn(\bolbeta)$, as shown on the plot.  The envelope represents the subspace in which the populations differ, which seems consistent with the pattern of variation shown in the plot.  The orthogonal complement of the envelope, represented by a dashed line on the plot, represents the immaterial variation.  The two populations are inferred to be the same when projected onto this subspace, which also seems consistent with the pattern of variation in the plot.  The envelope estimator of a mean difference is obtained by first projecting the points onto the envelope and thus removing the immaterial variation, and then projecting the points onto the horizontal or vertical axis.  The two density estimates with the smaller variation represent this operation.  These densities are well separated, leading to increased efficiency.

\begin{figure}
\begin{centering}
\includegraphics[bb = 130bp 240bp 480bp 500bp,scale=0.75]{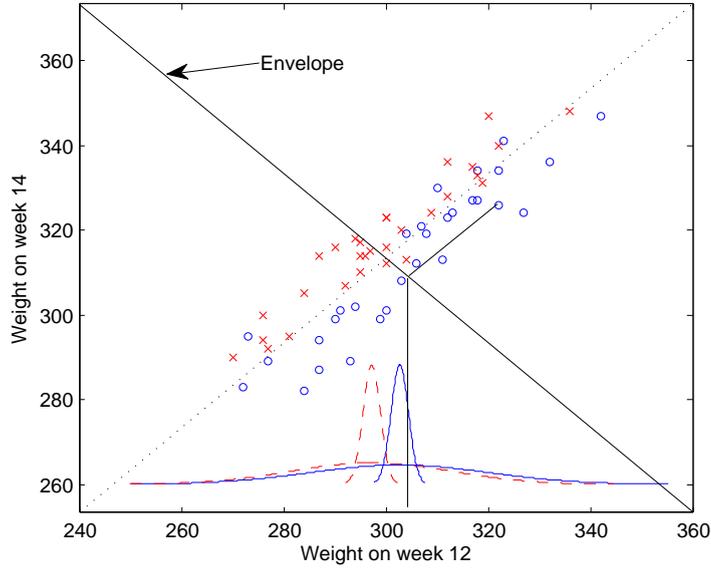}
\par\end{centering}
\caption{\label{fig:cow} Kenward's cow data with the 30 animals receiving one treatment marked as o's and the 30 animals receiving the other marked as x's.The curves on the bottom are  densities of $Y_6|(X=0)$ and $Y_6|(X=1)$: the flat two curves are obtained by projecting the data onto the $Y_6$ axis (standard analysis), and the two other densities are obtained by first project the data onto the envelope and then onto the $Y_6$ axis (envelope analysis).  One representative projection path is shown on the plot.}
\end{figure}

\begin{table}
\begin{centering}
\begin{tabular}{|c|c|c|c|c|c|c|c|c|c|c|}
\hline
\multicolumn{11}{|c|}{OLS estimator}\tabularnewline
\hline
Week  & 2 & 4 & 6 & 8 & 10 & 12 & 14 & 16 & 19 & 19\\
\hline
$\hatbolbeta_{\OLS}$  & 2.4 & 3.3 & 3.1 & 4.7 & 4.7 & 5.5 & -4.8 & -4.5 & -2.8 & 5.0
\tabularnewline
\hline
Bootstrap SE  & 2.9 & 3.2 & 3.5 & 3.6 & 4.0 & 4.2 & 4.4 & 4.5 & 5.4 & 6.0
\tabularnewline
\hline
\hline
\multicolumn{11}{|c|}{Envelope estimator}\tabularnewline
\hline
$\hatbolbeta_{\Env}$  & -2.2 & -0.5 &  0.9 &  2.4 & 2.9 & 5.4 & -5.1 & -4.6 & -3.7 & 4.2
\tabularnewline
\hline
Bootstrap SE  & 1.13 & 0.84 & 1.07 & 1.03 & 0.81 & 1.12 & 1.07 & 1.04 & 1.08 & 1.02\\
\hline
Asymptotic SE/$\sqrt{n}$  & 0.88 & 0.74 & 0.72 & 0.84 & 0.70 & 1.02 & 0.92 & 0.86 & 0.90 & 0.85
\tabularnewline
\hline
\hline
\multicolumn{11}{|c|}{Bootstrap SE ratios of OLS estimator over envelope estimator}\tabularnewline
\hline
SE ratios & 2.6 & 3.8 & 3.3 &3.5&5.0&3.7&4.1&4.3&5.0&5.9\tabularnewline
\hline
\end{tabular}
\par\end{centering}
\caption{\label{tab:cow}Bootstrap standard errors of the 10 elements in $\hatbolbeta$ under the OLS estimator and the envelope estimator with $u=1$. The bootstrap standard errors were estimated using 100 bootstrap samples.}
\end{table}

\section{Objective functions for estimating an envelope}\label{sec:objectiveFun}

\subsection{The objective function and its properties}

In this section we propose a generic objective function for estimating a basis $\bolGamma$ of an arbitrary envelope $\calE_{\mbM}(\calB) \subseteq \mbbR^{\d}$, where $\mbM\in\mbbS^{\d}$ is a symmetric positive definite matrix.
Let $\calB$ be spanned by a $\d\times \d$ matrix $\mbU$ so that $\calE_{\mbM}(\calB)=\calE_{\mbM}(\mbU)$.
Because $\spn(\mbU)=\spn(\mbU\mbU^T)$, we can always denote the envelope by $\calE_{\mbM}(\mbU)$ for some symmetric matrix $\mbU\geq0$.
We propose the following generic population objective function for estimating $\calE_{\mbM}(\mbU)$:
\begin{equation}\label{JG}
J(\bolGamma) = \log|\bolGamma^{T} \mbM \bolGamma| + \log| \bolGamma^{T}(\mbM+\mbU)^{-1}\bolGamma|,
\end{equation}
where $\bolGamma\in\mbbR^{\d\times u}$ denotes a semi-orthogonal basis for elements in Grassmann manifold $\calG_{u,\d}$,  $u$ is the dimension of the envelope, and $u < d$.  We refer to the operation of optimizing (\ref{JG}) or its sample version given later in (\ref{JnG}) as full Grassmann (\FG) optimization.
Since $J(\bolGamma)=J(\bolGamma\mbO)$ for any
orthogonal $u\times u$ matrix $\mbO$, the minimizer $\tilbolGamma=\arg\min_{\bolGamma}J(\bolGamma)$
is not unique. But we are interested only in $\spn(\tilbolGamma)$,
which is unique as shown in the following proposition.

\begin{pn} \label{pn: prop obj} Let $\tilbolGamma\in\mbbR^{\d\times u}$ be a minimizer of $J(\bolGamma)$. Then $\spn(\tilbolGamma)=\calE_{\mbM}(\mbU)$. \end{pn}

To gain intuition on how $J(\bolGamma)$ is minimized by any $\tilbolGamma$ that spans the envelope $\calE_{\mbM}(\mbU)$, we let $(\bolGamma,\bolGamma_0)\in\mbbR^{\d\times \d}$ be an orthogonal matrix and decompose the objective function into two parts: $J(\bolGamma)=J^{(1)}(\bolGamma)+J^{(2)}(\bolGamma)$, where
\begin{eqnarray*}
J^{(1)}(\bolGamma)
&=&
\log|\bolGamma^T\mbM\bolGamma| + \log|\bolGamma_0^T\mbM\bolGamma_0|,\\
J^{(2)}(\bolGamma)
&=&
\log|\bolGamma^T(\mbM+\mbU)^{-1}\bolGamma| - \log|\bolGamma_0^T\mbM\bolGamma_0|
\\
&=&
\log|\bolGamma_0^T(\mbM+\mbU)\bolGamma_0| - \log|\bolGamma_0^T\mbM\bolGamma_0|
- \log|\mbM+\mbU|.
\end{eqnarray*}
The first function $J^{(1)}(\bolGamma)$ is minimized by any $\bolGamma$ that spans a reducing subspace of $\mbM$.
Minimizing the second function $J^{(2)}(\bolGamma)$ is equivalent to minimizing $\log|\bolGamma_0^T(\mbM+\mbU)\bolGamma_0| - \log|\bolGamma_0^T\mbM\bolGamma_0|$, which is no less than zero and equals to zero when $\bolGamma_0^T\mbU\bolGamma_0=0$.
Thus $J^{(2)}(\bolGamma)$ is minimized by any $\bolGamma$ such that $\bolGamma_0^T\mbU\bolGamma_0=0$, or equivalently, $\spn(\mbU)\subseteq\spn(\bolGamma)$.
These properties of $J^{(1)}(\bolGamma)$ and $J^{(2)}(\bolGamma)$ are combined by $J(\bolGamma)$ to get a reducing subspace of $\mbM$ that contains $\spn(\mbU)$.
In the context of multivariate linear regression, minimizing $J^{(2)}(\bolGamma)$ is related to minimizing the residual sum of squares and minimizing $J^{(2)}$ is in effect pulling the solution towards principal components of responses or predictors.
Finally, because $u$ is the dimension of the envelope, the minimizer $\spn(\tilbolGamma)$ is unique by Definition~\ref{def: CLC env}.

The sample version $J_{n}$ of $J$ based on a sample of size $n$ is constructed by substituting estimators $\hatmbM$ and $\hatmbU$ of $\mbM$ and $\mbU$:
\begin{equation}\label{JnG}
J_n(\bolGamma) = \log|\bolGamma^{T} \hatmbM \bolGamma| + \log|\bolGamma^{T}(\hatmbM+\hatmbU)^{-1}\bolGamma|.
\end{equation}
Proposition~\ref{pn: prop obj} shows Fisher consistency of minimizers from optimizing the population objective function.
Furthermore, $\sqrt{n}$-consistency of $\hatbolGamma = \arg \min_{\bolGamma}J_{n}(\bolGamma)$ is stated in the following proposition.

\begin{pn} \label{pn: rootn consistent obj} Let $\widehat{\mbM}$ and $\widehat{\mbU}$ denote $\sqrt{n}$-consistent estimators for $\mbM > 0$ and $\mbU \geq 0$. Let $\hatbolGamma\in\mbbR^{\d\times u}$ be a minimizer of $J_n(\bolGamma)$, then $\mbP_{\hatbolGamma}$ is $\sqrt{n}$-consistent for the projection onto $\calE_{\mbM}(\mbU)$. \end{pn}

When we connect the objective function $J_n(\bolGamma)$ with multivariate linear models in Section~\ref{subsec:connection}, we will find that previous likelihood-based envelope objective functions can be written in form (\ref{JnG}).
The likelihood approach to envelope estimation is based on normality assumptions for the conditional distribution of the response given the predictors or the joint distribution of the predictors and responses.  The envelope objective function arising from this approach is a partially maximized log-likelihood obtained broadly a follows.
After incorporating the envelope structure into the model, partially maximize the normal log-likelihood function $L_n(\bolpsi,\bolGamma)$ over all the other parameters $\bolpsi$ with $\bolGamma$ fixed.
This leads to a likelihood-based objective function $L_n(\bolGamma)$, which equals a constant plus $-(n/2)J_n(\bolGamma)$ with $\hatmbM$ and $\hatmbU$ depending on context.
Proposition~\ref{pn: rootn consistent obj} indicates that the function $J_n(\bolGamma)$ can be used as a generic moment-based objective function requiring only $\sqrt{n}$-consistent matrices $\hatmbM$ and $\hatmbU$.
Consequently, normality is not a requirement for estimators based on $J_n(\bolGamma)$ to be useful, a conclusion that is supported by previous work and by our experience.
\FG\ optimization of $J_n(\bolGamma)$ can be computationally intensive and can require a good initial value. The \oneD\ algorithm in Section~\ref{sec:1Dalgorithm} mitigates the computational issues.

\subsection{Connections with previous work\label{subsec:connection}}
Envelope applications have so far been mostly restricted to the homoscedastic
multivariate linear model
\begin{equation}\label{mlm}
\mbY_{i} = \bolalpha + \bolbeta\mbX_{i} + \bolvarepsilon_{i}, \;i=1,\ldots,n,
\end{equation}where
$\mbY \in \mbbR^{r}$, the predictor vector $\mbX \in \mbbR^{p}$, $\bolbeta \in \mbbR^{r \times p}$, $\bolalpha \in \mbbR^{r}$ and the errors $\bolvarepsilon_{i}$ are independent copies of the normal random vector $\bolvarepsilon \sim N(0, \bolSigma)$.  The maximum likelihood estimators of $\bolbeta$ and $\bolSigma$ are then $\hatbolbeta_{\OLS} = \mbSYX \mbSX^{-1}$ and $\hatbolSigma= \mbRYX$.

\subsubsection{ Response envelopes}
Cook, et al. (2010) studied response envelopes for estimation of the coefficient matrix $\bolbeta$.
They conditioned on the observed values of $\mbX$ and motivated their developments by allowing for the possibility that some linear combinations of the response vector $\mbY$ are immaterial to the estimation of $\bolbeta$, as described previously in Section~\ref{sec:concepts}.  Reiterating,
suppose that there is an orthogonal matrix $(\bolGamma, \bolGamma_{0}) \in \mbbR^{r \times r}$ so that (i) $\spn(\bolbeta) \subseteq \spn(\bolGamma)$ and (ii) $\bolGamma^{T}\mbY \indep \bolGamma_{0}^{T}\mbY \mid \mbX$.
This implies that $(\mbY, \bolGamma^{T}\mbX) \indep \bolGamma_{0}^{T}\mbX$ and thus that $ \bolGamma_{0}^{T}\mbX$ is immaterial to the estimation of $\bolbeta$.
The smallest subspace $\spn(\bolGamma)$ for which these conditions hold is the $\bolSigma$-envelope of $\spn(\bolbeta)$, $\calE_{\bolSigma}(\bolbeta)$.

To determine the \FG\ estimator of $\calE_{\bolSigma}(\bolbeta)$, we let $\hatmbM = \mbRYX$ and $\hatmbM + \hatmbU = \mbSY$ in the objective function $J_n(\bolGamma)$ to reproduce the likelihood-based objective function in Cook et al. (2010). Then the maximum likelihood envelope estimators are $\hatbolbeta_{\Env} = \mbP_{\hatbolGamma}\hatbolbeta$ and $\hatbolSigma_{\Env} =  \mbP_{\hatbolGamma} \mbRYX  \mbP_{\hatbolGamma} + \mbQ_{\hatbolGamma} \mbRYX  \mbQ_{\hatbolGamma}$, where $\hatbolGamma=\arg\min J_n(\bolGamma)$.
Assuming normality for $\bolvarepsilon_i$, Cook et al. (2010) showed that the asymptotic variance of the envelope estimator $\hatbolbeta_{\Env}$ is no larger than that of the usual least squares estimator $\hatbolbeta$.
Under the weaker condition that $\bolvarepsilon_i$ are independent and identically distributed with finite fourth moments, the sample covariance matrices $\hatmbM$ and $\hatmbU$ are $\sqrt{n}$-consistent for $\mbM=\bolSigma$ and $\mbU=\bolSigma_{\mbY}-\bolSigma=\bolbeta\bolSigma^{-1}_{\mbX}\bolbeta^T$.
By Proposition~\ref{pn: rootn consistent obj}, we have $\sqrt{n}$-consistency of the envelope estimator $\hatbolbeta_{\Env}$ under this weaker condition.

\subsubsection{ Partial envelopes}
Su and Cook (2011) used the  $\bolSigma$-envelope of $\spn(\bolbeta_{1})$, $\calE_{\bolSigma}(\bolbeta_{1})$, to develop
a partial envelope estimator of $\bolbeta_{1}$
 in  the partitioned multivariate linear regression
\begin{equation}\label{pmlm}
\mbY_{i}=\bolalpha+\bolbeta\mbX_{i}+\bolvarepsilon_{i}
=\bolalpha+\bolbeta_{1}\mbX_{1i}+\bolbeta_{2}\mbX_{2i}+\bolvarepsilon_{i},\; i=1,\ldots,n,
\end{equation}
where $\bolbeta_{1}\in\mbbR^{r\times p_1}$, $p_1\leq p$, is the parameter vector of interest, $\mbX = (\mbX_{1}^T,\mbX_{2}^{T})^{T}$, $\bolbeta = (\bolbeta_{1},\bolbeta_{2})$ and the remaining terms are as defined for model (\ref{mlm}).
In this formulation, the immaterial information is $\bolGamma_{0}^{T}\mbY$, where $\bolGamma_{0}$ is a basis for $\calE_{\bolSigma}^{\perp}(\bolbeta_{1})$.
Since $\calE_{\bolSigma}(\bolbeta_{1}) \subseteq \calE_{\bolSigma}(\bolbeta)$, the partial envelope estimator $\hatbolbeta_{1,\Env} = \mbP_{\hatbolGamma}\hatbolbeta_{1}$  has the potential to yield efficiency gains beyond those for the full envelope, particularly when $\calE_{\bolSigma}(\bolbeta) = \mbbR^{r}$ so the full envelope offers no gain.  In the maximum likelihood estimation of $\bolGamma$, the same forms of $\hatmbM$, $\hatmbU$ and $J_n(\bolGamma)$ are used for partial envelopes $\calE_{\bolSigma}(\bolbeta_{1})$, except the roles of $\mbY$ and $\mbX$ in the usual response envelopes are replaced with the residuals: $\mbR_{\mbY|\mbX_{2}}$, residuals from the linear fits of $\mbY$ on $\mbX_{2}$, and $\mbR_{\mbX_{1}|\mbX_{2}}$, the residuals of $\mbX_{1}$ on $\mbX_{2}$.
Setting  $\hatmbM=\mbS_{\mbR_{\mbY|\mbX_{2}}|\mbR_{\mbX_1|\mbX_2}}=\mbS_{\mbY|\mbX}$ and $\hatmbM+\hatmbU=\mbS_{\mbY|\mbX_{2}}$ in the objective function $J_n(\bolGamma)$ reproduces the likelihood objective function of Su and Cook.  Again, Proposition~\ref{pn: rootn consistent obj} gives $\sqrt{n}$-consistency without normality.

\subsubsection{Predictor envelopes}\label{sec:predenvelopes}
Cook, et al. (2013) studied predictor reduction in model (\ref{mlm}), except the predictors are now stochastic with $\Var(\mbX) = \bolSigma_{\mbX}$ and $(\mbY, \mbX)$ was assumed to be normally distributed for the construction of maximum likelihood estimators.  Their reasoning, which parallels that for response envelopes, lead them to parameterize the linear model in terms of $\calE_{\bolSigma_{\mbX}}(\bolbeta^{T})$ and to achieve similar substantial gains in the estimation of $\bolbeta$ and in prediction.  The immaterial information in this setting is given by $\bolGamma_{0}^{T}\mbX$, where $\bolGamma_{0}$ is now a basis for $\calE^{\perp}_{\bolSigma_{\mbX}}(\bolbeta^{T})$.  They also showed that the SIMPLS algorithm for partial least squares provides a $\sqrt{n}$-consistent estimator of $\calE_{\bolSigma_{\mbX}}(\bolbeta^{T})$ and demonstrated that the envelope estimator $\hatbolbeta_{\Env} = \hatbolbeta\mbP_{\hatbolGamma(\mbS_{\mbX})}^{T}$ typically outperforms the SIMPLS estimator in practice.
For predictor reduction in model (\ref{mlm}), the envelope $\calE_{\bolSigma_{\mbX}}(\bolbeta^{T})$  is estimated with $\hatmbM = \mbRXY$, $\hatmbM + \hatmbU = \mbSX$.  As with response and partial envelopes, Proposition~\ref{pn: rootn consistent obj} gives us $\sqrt{n}$-consistency without requiring normality for $(\mbY, \mbX)$.

Techniques for estimating the dimension of an envelope are discussed in the parent articles of these methods, including use of an information criterion like BIC, cross validation or a hold-out sample.

\subsection{New envelope estimators inspired by the objective function}\label{sec:new}

The objective function $J_n(\bolGamma)$ can also be used for envelope estimation in new problems.
For example, to estimate the multivariate mean $\bolmu\in\mbbR^r$ in the model $\mbY = \bolmu + \bolvarepsilon$,
we can use the $\bolSigma$-envelope of $\spn(\bolmu)$ by taking
$\mbM=\bolSigma$ and $\mbU=\bolmu\bolmu^T$, whose sample versions are: $\hatmbM = \mbSY$, $\hatmbU = \hatbolmu\hatbolmu^{T}$ and  $\hatbolmu = n^{-1}\sum_{i=1}^{n}\mbY_{i}$.
Then substituting $\hatmbM$ and $\hatmbU$ leads to the same objective function $J_n(\bolGamma)$ as that obtained when deriving the likelihood-based envelope estimator from scratch.

For the second example, let $\mbY_i\sim N_r(\bolmu,\bolSigma)$, $i=1,\dots,n$, consist of longitudinal measurements of $n$ subjects over $r$ fixed time points.
Suppose we are not interested in the overall mean $\bar{\mu}=\mathbf{1}_r^T\bolmu/r\in\mbbR^1$ but rather interest centers on the deviations at each time point $\bolalpha=\bolmu-\bar{\mu}\mathbf{1}_r\in\mbbR^r$.
Let $\mbQ_{\mathbf{1}}=\mbI_r-\mathbf{1}_r\mathbf{1}_r^T/r$ denote the projection onto the orthogonal complement of $\spn(\mathbf{1}_r)$.
Then $\bolalpha=\mbQ_{\mathbf{1}}\bolmu$ and we consider estimating the constrained envelope: $\calE_{\mbQ_{\mathbf{1}}\bolSigma\mbQ_{\mathbf{1}}}(\mbQ_{\mathbf{1}}\bolmu\bolmu^T\mbQ_{\mathbf{1}}):=\calE_{\mbM}(\mbU)$.
Optimizing $J_n(\bolGamma)$ with $\hatmbM = \mbQ_{\mathbf{1}}\mbSY\mbQ_{\mathbf{1}}$ and $\hatmbU = \mbQ_{\mathbf{1}}\hatbolmu\hatbolmu^{T}\mbQ_{\mathbf{1}}$ will again lead to the maximum likelihood estimator and to $\sqrt{n}$-consistency without normality.
Later from Proposition~\ref{pn: seq breakdown envelope}, we will see that $\calE_{\mbM}(\mbU)=\mbQ_{\mathbf{1}}\calE_{\bolSigma}(\bolmu\bolmu^T)$ and the optimization can be simplified.

The objective function $J_n(\bolGamma)$ introduces also a way of extending envelope regression semi-parametrically or non-parametrically.
This can be done by simply replacing the sample covariances $\hatmbM$ and $\hatmbU$ in Section~\ref{subsec:connection} with their semi-parametric and non-parametric counterparts.
Given a multivariate model $\mbY=\mbf(\mbX)+\bolepsilon$, where $\bolbeta\in\mbbR^p$, $\mbY\in\mbbR^r$ and $\mbf(\cdot):\ \mbbR^p\rightarrow\mbbR^r$, the envelope for reducing the response can be estimated by taking   $\hatmbM$ equal to the sample covariance of the residuals: $\hatmbM = n^{-1}\sum_{i=1}^{n}\{\mbY_i-\widehat{\mbf}(\mbX_i)\}
\{\mbY_i-\widehat{\mbf}(\mbX_i)\}^T$, and $\hatmbM+\hatmbU=\mbSY$.

\section{A \oneD\ algorithm \label{sec:1Dalgorithm}}

In this section we propose a method for estimating a basis $\bolGamma$ of an arbitrary envelope  $\calE_{\mbM}(\calB) \subseteq \mbbR^{\d}$ based on a series of one-dimensional optimizations.  The resulting algorithm is fast and stable, does not require carefully chosen starting values and the estimator it produces converges at the root-$n$ rate. The estimator can be used as it stands, or as a $\sqrt{n}$-consistent starting value for (\ref{JnG}).
In the latter case, one Newton-Raphson step from the starting value provides an estimator that is asymptotically equivalent under normality to the maximum likelihood estimators discussed in Section~\ref{subsec:connection} (Lehmann and Casella, 1998, p. 454.)
As mentioned in the Introduction, the algorithm we present here is an extension to general problems of the one-dimensional algorithm of Cook and Zhang (2014).

The population algorithm described in this section extracts one dimension at a time from $\calE_{\mbM}(\calB)=\calE_{\mbM}(\mbU)$ until a basis is obtained. It requires only $\mbM>0$, $\mbU\geq0$ and $u = \dim(\calE_{\mbM}(\calB))$ as previously defined in Section~\ref{sec:objectiveFun}.
Sample versions are obtained by substituting $\sqrt{n}$-consistent estimators $\hatmbM$ and $\hatmbU$ for $\mbM$ and  $\mbU$.
Otherwise, the algorithm itself does not depend on a statistical context, although the manner in which the estimated basis is used subsequently does.

The following proposition is the basis for a sequential breakdown of a $u$-dimensional \FG\ optimization (see also Cook and Zhang (2014; Lemma 5)).

\begin{pn}\label{pn: seq breakdown envelope}
Let  $(\mbB,\mbB_{0})$ denote an orthogonal basis of $\mbbR^{\d}$,
where $\mbB\in\mbbR^{\d\times q}$, $\mbB_{0}\in\mbbR^{\d\times(\d-q)}$
and $\spn(\mbB)\subseteq\calE_{\mbM}(\calB)$. Then $\mbv\in\calE_{\mbB_{0}^{T}\mbM\mbB_{0}}(\mbB_{0}^{T}\calB)$
implies that $\mbB_{0}\mbv\in\calE_{\mbM}(\calB)$.
\end{pn}
Suppose we know an orthogonal basis $\mbB$ for a subspace of the envelope
$\calE_{\mbM}(\calB)$. Then by Proposition~\ref{pn: seq breakdown envelope} we can find the rest of $\calE_{\mbM}(\calB)$ by looking into $\calE_{\mbB_{0}^{T}\mbM\mbB_{0}}(\mbB_{0}^{T}\calB)$,
which is a lower dimensional envelope.
This then provides a motivation for Algorithm~\ref{alg: 1D manifold}, which sequentially constructs vectors $\mbg_{k} \in \calE_{\mbM}(\calB)$, $k=1,\dots,u$, until a basis is obtained, $\spn(\mbg_{1},\dots,\mbg_{u}) = \calE_{\mbM}(\calB)$.
This algorithm follows the structure implied by Proposition~\ref{pn: seq breakdown envelope} and the stepwise objective functions $J_{k}$ are each one-dimensional versions of (\ref{JG}).
The first direction $\mbg_{1}$ requires optimization in $\mbbR^{d}$, while the optimization dimension is reduced by 1 in each subsequent step.

\begin{algorithm}
\begin{enumerate}
\item Set initial value $\mbg_{0}=\mbG_{0}=0$.
\item For $k=0,\dots,u-1$,
\begin{enumerate}
\item Let $\mbG_{k}=(\mbg_{1},\dots,\mbg_{k})$ if $k\geq1$ and let $(\mbG_{k},\mbG_{0k})$ be an orthogonal basis for $\mbbR^{\d}$.
\item Define the stepwise objective function
    \begin{equation}\label{Jkw}
    \J_{k}(\mbw)=\log(\mbw^{T}\mbM_{k}\mbw)
    +\log\{\mbw^{T}(\mbM_{k}+\mbU_{k})^{-1}\mbw\},
    \end{equation}
    where $\mbM_{k}=\mbG_{0k}^{T}\mbM\mbG_{0k}$, $\mbU_{k}=\mbG_{0k}^{T}\mbU\mbG_{0k}$ and $\mbw\in\mbbR^{\d-k}$.
\item Solve $\mbw_{k+1}=\arg\min_{\mbw}\J_{k}(\mbw)$ subject to a length constraint $\mbw^T\mbw=1$.
\item Define $\mbg_{k+1}=\mbG_{0k}\mbw_{k+1}$ to be the unit length $(k+1)$-th stepwise direction.
\end{enumerate}
\end{enumerate}
\caption{\label{alg: 1D manifold} The \oneD\ algorithm.}
\end{algorithm}

\medskip
\noindent\textbf{Remark 1.} At step 2(c) of Algorithm~\ref{alg: 1D manifold}, we need to minimize the stepwise objective function $\J_k(\mbw)$ under the constraint that $\mbw^T\mbw=1$.
The {\tt sg\_min} package can still be used to deal with this constraint since we are optimizing over one-dimensional Grassmann manifolds.
An alternative way is to integrate the constraint $\mbw^T\mbw=1$ into the objective function in (\ref{Jkw}), so that we only need to minimize the unconstrained function
\begin{equation}\label{Jkw uncon}
\widetilde{\J}_k(\mbw)
=\log(\mbw^{T}\mbM_{k}\mbw)
+\log\{\mbw^{T}(\mbM_{k}+\mbU_{k})^{-1}\mbw\}
-2\log(\mbw^T\mbw),
\end{equation}
with an additional normalization step for its minimizer $\mbw_{k+1}\leftarrow\mbw_{k+1}/||\mbw_{k+1}||$.
This unconstrained objective function $\J_k(\mbw)$ can be solved by any standard numerical methods such as conjugate gradient or Newton's method.
We have implemented this idea with the general purpose optimization function {\tt optim} in R and obtained good results.

\medskip
\noindent\textbf{Remark 2.} We have also considered other types of sequential optimization methods for envelope estimation. For example, we considered minimizing $D_1(\mbw)$ at each step under orthogonality constraints such as $\mbw_{k+1}^T\mbw_{j}=0$ or $\mbw_{k+1}^T\mbM\mbw_{j}=0$ for $j\leq k$. These types of orthogonality constraints are used widely in PLS algorithms and principal components analysis. We find the statistical properties of these sequential methods are inferior to those of the \oneD\ algorithm.  For instance, they are clearly inferior in simulations and we doubt that they lead to consistent estimators.
\medskip

The next two propositions establish the Fisher consistency of Algorithm~\ref{alg: 1D manifold} in the population and the $\sqrt{n}$-consistency of its sample version.

\begin{pn} \label{pn: Fisher_consistent 1D manifold algorithm}
Assume that $\mbM>0$, and let $\mbG_u$ denote the end result of the algorithm.  Then $\spn(\mbG_u) = \calE_{\mbM}(\calB)$.
\end{pn}
\begin{pn} \label{pn: sqrt_n_consistent 1D manifold algorithm}
Assume that  $\mbM>0$ and let $\widehat{\mbM} > 0$ and $\widehat{\mbU}$
denote $\sqrt{n}$-consistent  estimators for $\mbM$ and $\mbU$.
Let $\widehat{\mbG}_{u}$ denote the estimator obtained from the \oneD\ algorithm using $\widehat{\mbM}$ and $\widehat{\mbU}$ instead of $\mbM$ and $\mbU$. Then $\mbP_{\widehat{\mbG}_{u}}$ is $\sqrt{n}$-consistent for the projection onto $\calE_{\mbM}(\calB)$.
\end{pn}
The algorithm discussed in this section can be used straightforwardly in the contexts of the three envelopes reviewed in Section~\ref{subsec:connection} and the extensions sketched in Section~\ref{sec:new}.
The statistical properties of the \oneD\ algorithm estimator stated in Propositions~\ref{pn: Fisher_consistent 1D manifold algorithm} and \ref{pn: sqrt_n_consistent 1D manifold algorithm} are exactly parallel to the properties of  \FG\  optimization in Propositions~\ref{pn: prop obj} and \ref{pn: rootn consistent obj}.

%

\section{Simulations}\label{sec:data}
In this section, we compare the \oneD\ algorithm to \FG\ (full Grassmann manifold) optimization, focusing on computational cost.
For fair comparisons, the implementation of our \oneD\ algorithm was based on minimizing the length-constrained objective function (\ref{Jkw}) using the {\tt sg\_min} package.
Implementation of the \oneD\ algorithm with other computing packages using the unconstrained objective function (\ref{Jkw uncon}) may offer even faster estimation procedures.

\subsection{Simulations}
We considered the response envelope model in Cook et al. (2010) with univariate predictor $X\sim N(0,1)$ and multivariate response $\mbY=\bolalpha + \bolbeta X + \bolepsilon$, where $\bolepsilon\sim N_r(0,\bolSigma)$ and we were interested in estimation of $\calE_{\bolSigma}(\bolbeta)$.
We generated $\mbM=\bolSigma$ and $\mbU=\bolbeta\bolbeta^T$ in accordance with an envelope structure: $\bolbeta=\bolGamma\boleta$ and $\bolSigma
=\bolGamma\bolOmega\bolGamma^{T}
+\bolGamma_{0}\bolOmega_{0}\bolGamma_{0}^{T}$ for some positive definite matrices $\bolOmega\in\mbbS^{u}$ and $\bolOmegao\in\mbbS^{r-u}$ and a vector of ones $\boleta=\mathbf{1}_{u}\in\mbbR^u$.
The semi-orthogonal basis $\bolGamma\in\mbbR^{r\times u}$
for $\calE_{\mbM}(\mbU)$ was randomly generated and $\bolGamma_0$ was then obtained so that $(\bolGamma,\bolGamma_{0})$ was an orthogonal basis for $\mbbR^{r}$.
The two covariance matrices $\bolOmega$, $\bolOmega_{0}$ were generated as $\mbA\mbA^{T}>0$, where $\mbA$ was a square matrix with corresponding dimensions and was filled with uniform $(0,1)$ random numbers.

We first examined the performances of our \oneD\ algorithm in the population. We generated 100 pairs of $\mbM$ and $\mbU$ for each of three dimension configurations, $(r, u)=(10, 3)$, $(r, u)=(30, 10)$ and $(r, u) = (70, 20)$.  These dimensions correspond to the real optimization dimensions $u(r-u) = 21$, $200$ and $1000$ for \FG\ optimization, while the 1D algorithm optimizes over at most $r-1$ real dimensions at each iteration.
We recorded the CPU time $T$ for estimating an envelope and the Frobenius  norm between the true envelope and an estimated envelope defined as $\mathrm{dist(\bolGamma,\tilbolGamma)}
=||\bolGamma\bolGamma^{T}-\tilbolGamma\tilbolGamma^{T}||_{F}$.
The results for running the \oneD\ algorithm (Algorithm \ref{alg: 1D manifold}) and the \FG\ optimization of (\ref{JnG})
are given  in the first three rows of Table~\ref{tab:computation time}.
Apparently the \oneD\ algorithm achieved the same accuracy as \FG\ optimization and was much less time-consuming, especially at the large dimension $(r,u)=(30,10)$ and $(r,u)=(70,20)$.

We next generated 100 replicated data sets for one pairs of $\mbM$ and $\mbU$, and used the sample estimator $\hatmbM=\mbS_{\mbY|X}$ and $\hatmbU=\mbS_{\mbY} - \mbS_{\mbY|X}$ for envelope estimation.
We let $n=400$ and kept the same dimensions.
From Table~\ref{tab:computation time}, we can see the \oneD\ algorithm outperformed \FG\ optimization in terms of computational efficiency.

For \FG\ optimization, we chose initial value according to the approach described in Su and Cook (2011; Section 3.5), first optimizing the objective function over the $2r$ eigenvectors of $\hatmbM$ and $\hatmbM+\hatmbU$.
This initial value search procedure alone could be computationally costly, but we did not include the time spent on this when we summarized the computing time $T$ for the \FG\ optimization algorithm in Table~\ref{tab:computation time}.  Additionally, we used only the true value of $u$ in each simulation.  The performance of optimizations at other than the true value of $u$, as necessary in the  application of BIC, need not follow those of Table~\ref{tab:computation time}, as we illustrate in the next section.

\begin{table}
\begin{centering}
\begin{tabular}{|c|c|c|c|c|}
\hline
 & \multicolumn{2}{c|}{\oneD\ algorithm} & \multicolumn{2}{c|}{FG optimization}\tabularnewline
\hline
 & $T$ & ${ \mathrm{dist}(\bolGamma,\hatbolGamma)}$ & $T$ & ${ \mathrm{dist}(\bolGamma,\hatbolGamma)}$\tabularnewline
\hline
\hline
$(n,r,u)=(\infty,10,3)$ & 2.0 (0.2) & $<1.0\times 10^{-8}$ & 6.6 (0.3) & $<1.0\times 10^{-8}$\tabularnewline
\hline
$(n,r,u)=(\infty,30,10)$ & 2.6 (0.1) & $<1.0\times 10^{-4}$ & 127 (11) & $<1.0\times 10^{-4}$\tabularnewline
\hline
$(n,r,u)=(\infty,70,20)$ & 447 (11) & $<1.0\times 10^{-2}$ & 5084 (1283) & $<1.0\times 10^{-2}$\tabularnewline
\hline
$(n,r,u)=(400,10,3)$ & 0.6 (0.04) & 1.1 (0.05) & 1.2 (0.09) & 1.0 (0.05)\tabularnewline
\hline
$(n,r,u)=(400,30,10)$ & 30.7 (0.6) & 2.8 (0.02) & 121 (7) & 3.1 (0.02)\tabularnewline
\hline
$(n,r,u)=(400,70,20)$ & 534 (5) & 4.6 (0.04) & 4187 (68) & 4.7 (0.03)\tabularnewline
\hline
\end{tabular}
\par\end{centering}

\caption{\label{tab:computation time} Comparisons between the \oneD\ algorithm and \FG\ optimization. Each cell contains the average running time in seconds over 100 simulations, with its standard error given in parentheses. The population algorithms with $\mbM$ and $\mbU$ were indicated with $n=\infty$ and the sample algorithms had $n=400$. }
\end{table}

\subsection{Starting values}

As mentioned previously, good starting values can be crucial to the performance of \FG\ optimization.  To highlight this point, we used the meat data analyzed previously by Cook et al. (2013) for envelope predictor reduction in multivariate linear regression.
This data set consists of spectral measurements from infrared transmittance for fat, protein and water for 103 meat samples.
Following Cook et al. (2013), we used the protein percentage as the univariate response.
The $p=50$ predictors were spectral measurements at every fourth wavelength between 850nm and 1050nm.
Using five-fold cross-validation prediction error as their criterion and $u$ varying from $1$ to $25$,  Cook et al. (2013) compared the \FG\ envelope estimator described in Section~\ref{sec:predenvelopes}  to the OLS and SIMPLS estimators.  The starting value for the \FG\ envelope estimator was the SIMPLS estimator, which is  $\sqrt{n}$-consistent in the context of predictor envelopes and had better performance than OLS.  SIMPLS was designed specifically for predictor reduction and is not applicable to response or partial reduction or to the extensions discussed in Section~\ref{sec:new}.  Their results showed the envelope estimator to be uniformly superior to OLS,  superior to SIMPLS for small values of $u$ and about the same as SIMPLS for large values of $u$.
In this study we used the same setup as Cook et al. (2013), except we focused on comparisons between the \oneD\ algorithm and  the \FG\  envelope estimator with starting values again chosen following the approach described in Su and Cook (2011; Section 3.5), since the  $2r$ eigenvectors of $\hatmbM$ and $\hatmbM+\hatmbU$ may be all that is easily available without recourse to the 1D algorithm.

We plotted in Figure~\ref{fig: Meat 1D vs MLE} (top two plots) the five-fold cross-validation squared prediction error and the elapsed CPU time (in seconds) for computing the \FG\ envelope estimators with dimensions $u=1,\dots,25$.
Although we had five-folds and thus estimated five envelopes for each dimension $u$,  the time reported is the average for estimating one envelope.  The number of real optimization dimensions $u(50-u)$ varied between $49$ and $625$.
For the larger values of $u$,  \FG\ optimization took a very long time to compute, so we capped the number of allowed iterations at 5000.
For small dimensions, $u\leq3$,  \FG\ optimization and the \oneD\ algorithm had close prediction performance, and there were no convergence issues.
For $u=4$ and $5$, \FG\ optimizations tended to become trapped into local minima, as indicated by the prediction error. For larger dimensions, $u>10$, \FG\ optimization began bumping into the iteration limit.
The computation time for the \oneD\ method was almost linearly increasing in $u$ because of the sequential manner of the algorithm. With increasing number of components, the prediction errors of both methods converged towards that of the ordinary least squares estimator as expected, since they both reduce to ordinary least squares when $u = 50$.
However, the \oneD\ algorithm provided better estimators, consistently over $u$, than the OLS estimator and the \FG\ envelope estimator.

This difference in the results reported by Cook et al. (2013) and the results shown in the top plot of Figure~\ref{fig: Meat 1D vs MLE} arises because of the different staring values.
In Cook et al. (2013), the initial values were  $\sqrt{n}$-consistent, while here  we chose initial values from the eigenvectors of $\hatmbM$ and $\hatmbM+\hatmbU$.
When using these starting values, \FG\ optimizations tended to get trapped by local minima that were close to the initial values, which accounts for the inferior performance of the \FG\ envelope estimator in this setting.
From Lehmann and Casella (1998; Theorem 4.3), we know that one Newton-Raphson iteration from any $\sqrt{n}$-consistent estimator, the \oneD\ algorithm estimator for instance, will be asymptotically equivalent to the MLE, even if there were local minima.
We used 100 iterations (instead of one) for the \FG\ optimization with \oneD\ algorithm estimators as initial values.
The cross-validation prediction errors, shown in the bottom plot of Figure~\ref{fig: Meat 1D vs MLE}, were very close to those of the 1D algorithm.  The \FG\ algorithm did a little bit worse than the 1D algorithm at some $u$ because with 100 iterations it occasionally got trapped in a local minimum as it tried to improve the starting value.

\begin{figure}
\begin{centering}
\vspace{1.3in}
\includegraphics[bb = 130bp 180bp 480bp 460bp,scale=0.9]{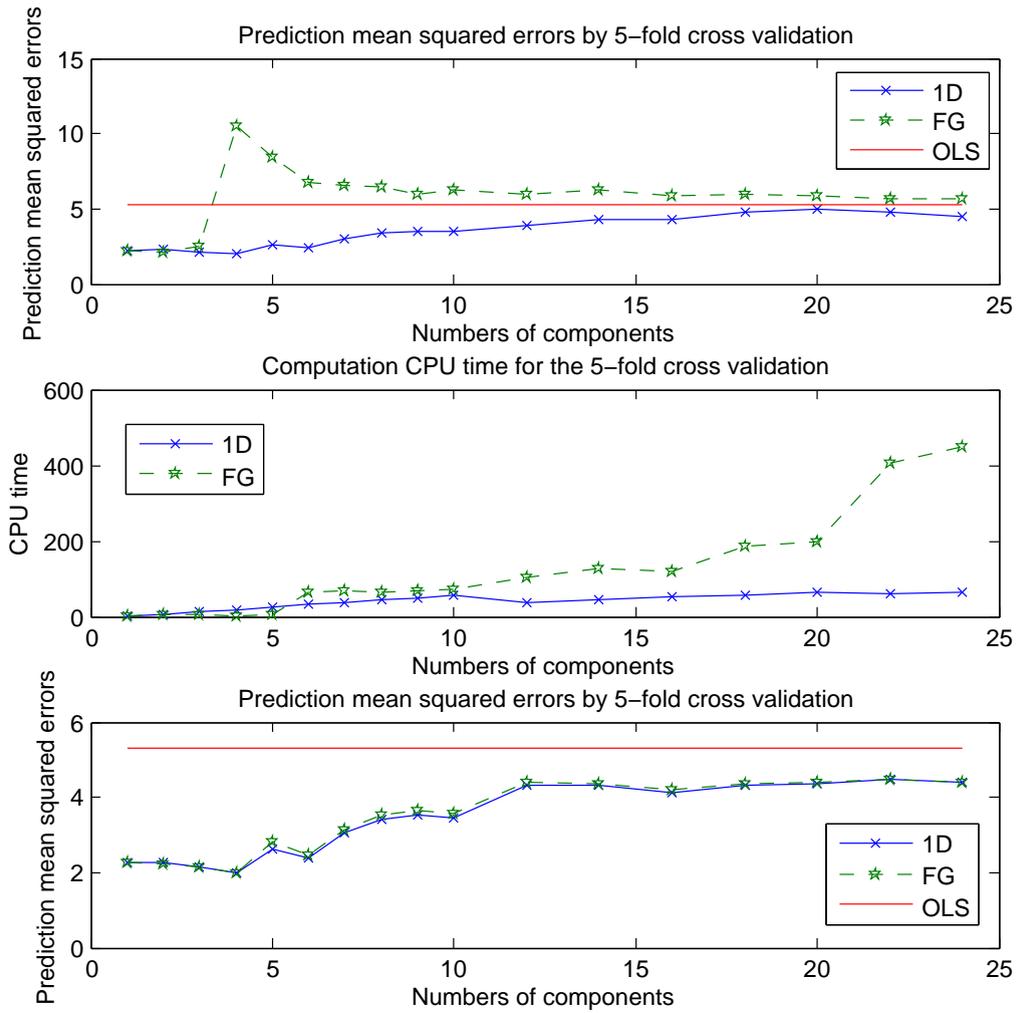}
\par\end{centering}
\caption{\label{fig: Meat 1D vs MLE} Meat protein data. The \FG\ optimizations shown in the top and the middle plots were based on starting values suggested by Cook and Su (2011; Section 5.3). And the \FG\ optimization in bottom plot was using the \oneD\ algorithm estimators as starting value.}
\end{figure}


\section{Conclusion}\label{sec:conclusion}

Our study led to the following conclusions.
The \FG\ envelope estimator (\ref{JnG}) can be computed straightforwardly when the number of real dimensions $u(k-u)$ is relatively small, say less than $150$, as illustrated in the example of Section~\ref{sec:concepts}.  When this dimension is large, computing time and local minima can become serious issues, and then root-$n$ consistent starting values become crucial.  The 1D algorithm can be used confidently for starting values, or as a stand-alone algorithm for envelope estimation.

\section*{Acknowledgments}
Research for this article was supported in part by grant DMS-1007547 from the National Science Foundation.

\appendix
\global\long\def\thesection{\Alph{section}}
\setcounter{equation}{0}
\renewcommand{\theequation}{A\arabic{equation}}
\setcounter{pn}{0}
\renewcommand{\thepn}{A\arabic{pn}}
\begin{appendix} 

\section{Appendix: Proofs and Technical Details}
\subsection{Proposition~\ref{pn: prop obj}}
The proof of this proposition is very similar to the proof of Proposition 4.2 in Cook et al. (2013), thus is omitted.

\subsection{Proposition~\ref{pn: rootn consistent obj}}
The proof follows from Proposition~\ref{pn: Amemiya 4.1.1} and Proposition~\ref{pn: Amemiya 4.1.3} in the same way as Proposition~\ref{pn: sqrt_n_consistent 1D manifold algorithm} in Section~\ref{supp subsec: root-n 1D}. Thus we omit the details of the proof.

\subsection{Proposition \ref{pn: seq breakdown envelope}}

\begin{proof}

From our set-up, we know that $\mbB^{T}\mbM\mbB>0$ thus $\calE_{\mbB^{T}\mbM\mbB}(\mbB^{T}\calB)$
exists. Let $\bolGamma$ be a basis of $\calE_{\mbM}(\calB)$, and
$(\bolGamma,\bolGammao)$ be a orthogonal basis of $\mbbR^{p}$, then
$\mbM=\bolGamma\bolOmega\bolGamma^{T}+\bolGammao\bolOmegao\bolGamma_{0}^{T}$
and $\calB\subseteq\spn(\bolGamma)$ for some symmetric matrices $\bolOmega>0$
and $\bolOmegao>0$. Therefore,
\begin{eqnarray}
\mbB_{0}^{T}\mbM\mbB_{0} & = & (\mbB_{0}^{T}\bolGamma)\bolOmega(\mbB_{0}^{T}\bolGamma)^{T}+(\mbB_{0}^{T}\bolGamma_{0})\bolOmegao(\mbB_{0}^{T}\bolGamma_{0})^{T}\nonumber \\
\mbB_{0}^{T}\calB & \subseteq & \spn(\mbB_{0}^{T}\bolGamma),\label{B0calB}
\end{eqnarray}
where $\spn(\mbB_{0}^{T}\bolGamma)$ is the orthogonal compliment
of $\spn(\mbB_{0}^{T}\bolGammao)$ in $\mbbR^{p-q}$ since $\spn(\mbB)\subseteq\spn(\bolGamma)$.
Then we see that
\begin{equation}
\mbB_{0}^{T}\mbM\mbB_{0}=\mbP_{\mbB_{0}^{T}\bolGamma}\mbB_{0}^{T}\mbM\mbB_{0}\mbP_{\mbB_{0}^{T}\bolGamma}+\mbQ_{\mbB_{0}^{T}\bolGamma}\mbB_{0}^{T}\mbM\mbB_{0}\mbQ_{\mbB_{0}^{T}\bolGamma},
\end{equation}
which implies that $\spn(\mbB_{0}^{T}\bolGamma)$ is a reducing subspace
of $\mbB_{0}^{T}\mbM\mbB_{0}$ which also contains $\mbB_{0}^{T}\calB$
by (\ref{B0calB}). By definition, we know that $\calE_{\mbB_{0}^{T}\mbM\mbB_{0}}(\mbB_{0}^{T}\calB)$
is the smallest reducing subspace of $\mbB_{0}^{T}\mbM\mbB_{0}$ that
contains $\mbB_{0}^{T}\calB$. Hence $\calE_{\mbB_{0}^{T}\mbM\mbB_{0}}(\mbB_{0}^{T}\calB)\subseteq\spn(\mbB_{0}^{T}\bolGamma)$.
Thus $\mbv\in\calE_{\mbB_{0}^{T}\mbM\mbB_{0}}(\mbB_{0}^{T}\calB)$
implies $\mbB_{0}\mbv\in\calE_{\mbM}(\calB)$.

\end{proof}

\subsection{Proposition \ref{pn: Fisher_consistent 1D manifold algorithm}}

\begin{proof}
We first write
\begin{eqnarray*}
\mbM & = & \bolGamma\bolPhi\bolGamma^{T}+\bolGamma_{0}\bolOmegao\bolGamma_{0}^{T},\\
\mbM+\mbU & = & \bolGamma\bolOmega\bolGamma^{T}+\bolGammao\bolOmegao\bolGamma_{0}^{T},
\end{eqnarray*}
where $\bolOmega_{0}>0$, $\bolOmega>0$, $\bolPhi>0$, $\bolOmega-\bolPhi\geq0$,
$\bolGamma$ is semi-orthogonal basis for $\calE_{\mbM}(\calB)$ and
$(\bolGamma,\bolGamma_{0})\in\mbbR^{p}$ is orthogonal basis for $\mbbR^{p}$.

We begin by considering optimization for the first direction $\mbg_{1}=\arg\min_{\mbg\in\mbbR^{p}}J_0(\mbg),$
where $J_0(\mbg)=\log|\mbg^{T}\mbM\mbg|+\log|\mbg^{T}(\mbM+\mbU)^{-1}\mbg|$
and the minimization is subject to the constraint $\mbg^{T}\mbg=1$. Let
$\mbg=\bolGamma\mbh+\bolGammao\mbho$ for some $\mbh\in\mbbR^{u}$
and $\mbho\in\mbbR^{(p-u)}$. Consider the optimization problem as
the unconstrained problem,
\[
\mbg_{1}=\arg\min_{\mbg\in\mbbR^{p}}\left\{ \log|\mbg^{T}\mbM\mbg|+\log|\mbg^{T}(\mbM+\mbU)^{-1}\mbg|-2\log|\mbg^{T}\mbg|\right\} .
\]
Then we will have the same solution as the original problem up to
an arbitrary scaling constant. Next, we plug-in these expressions
for $\mbg$, $\mbM+\mbU$ and $\mbM$,
\begin{eqnarray*}
 &  & \log|\mbg^{T}\mbM\mbg|+\log|\mbg^{T}(\mbM+\mbU)^{-1}\mbg|-2\log|\mbg^{T}\mbg|\\
 & = & \log\{\mbh^{T}\bolPhi\mbh+\mbh_{0}^{T}\bolOmegao\mbho\}+\log\{\mbh^{T}\bolOmega^{-1}\mbh+\mbh_{0}^{T}\bolOmega_{0}^{-1}\mbho\}-2\log\{\mbh^{T}\mbh+\mbh_{0}^{T}\mbho\}\\
 & \equiv & f(\mbh,\mbho).
\end{eqnarray*}
Taking partial derivative with respect to $\mbho$, we
have
\begin{eqnarray*}
\frac{\partial}{\partial\mbho}f(\mbh,\mbho) & = & \frac{2\bolOmegao\mbho}{\mbh^{T}\bolPhi\mbh+\mbh_{0}^{T}\bolOmegao\mbho}+\frac{2\bolOmega_{0}^{-1}\mbho}{\mbh^{T}\bolOmega^{-1}\mbh+\mbh_{0}^{T}\bolOmega_{0}^{-1}\mbho}-\frac{4\mbh_{0}}{\mbh^{T}\mbh+\mbh_{0}^{T}\mbho}.
\end{eqnarray*}
To get local minimums we need to set $\frac{\partial}{\partial\mbho}f(\mbh,\mbho)=0$
which gives the following equality.
\begin{eqnarray*}
\left\{ \frac{2\bolOmegao}{\mbh^{T}\bolPhi\mbh+\mbh_{0}^{T}\bolOmegao\mbho}+\frac{2\bolOmega_{0}^{-1}}{\mbh^{T}\bolOmega^{-1}\mbh+\mbh_{0}^{T}\bolOmega_{0}^{-1}\mbho}\right\} \mbho & = & \left\{ \frac{4}{\mbh^{T}\mbh+\mbh_{0}^{T}\mbho}\right\} \mbh_{0}.
\end{eqnarray*}
Define
\begin{eqnarray}
\mbA_{0} & = & \left\{ \frac{2\bolOmegao}{\mbh^{T}\bolPhi\mbh+\mbh_{0}^{T}\bolOmegao\mbho}+\frac{2\bolOmega_{0}^{-1}}{\mbh^{T}\bolOmega^{-1}\mbh+\mbh_{0}^{T}\bolOmega_{0}^{-1}\mbho}\right\} /\left\{ \frac{4}{\mbh^{T}\mbh+\mbh_{0}^{T}\mbho}\right\}.
\end{eqnarray}

Since $\bolOmega_{0}>0$, we know $\mbA_{0}>0$.
Then $\mbA_{0}\mbh_{0}=\mbh_{0}$
has solutions only as eigenvectors of $\mbA_{0}$. The eigenvectors of $\mbA_{0}$ are
 the same as those of $\bolOmegao$. Hence,  $\mbh_{0}$ equals
$0$ or any eigenvector $\ell_{k}(\bolOmega_{0})$ of $\bolOmega_{0}$.
Therefore, the minimum value of $f(\mbh,\mbho)$ has to be obtained
by $0$ or $\ell_{k}(\bolOmega_{0})$ (since $\mbh_{0}=\infty$ can
be easily eliminated). If $\mbho=0$ then our conclusion follows.

Assume $\mbho\neq0$ and $\bolOmegao\mbho=\lambda_{k}\mbho$. Then,
\begin{eqnarray*}
f(\mbh,\mbho) & = & \log\{\frac{\mbh^{T}\bolPhi\mbh+\lambda_{k}\mbh_{0}^{T}\mbho}{\mbh^{T}\mbh+\mbh_{0}^{T}\mbho}\}+\log\{\frac{\mbh^{T}\bolOmega^{-1}\mbh+\frac{1}{\lambda_{k}}\mbh_{0}^{T}\mbho}{\mbh^{T}\mbh+\mbh_{0}^{T}\mbho}\}\\
 & = & \log\{\frac{\mbh^{T}\bolPhi\mbh}{\mbh^{T}\mbh}W_{h}+\lambda_{k}(1-W_{h})\}+\log\{\frac{\mbh^{T}\bolOmega^{-1}\mbh}{\mbh^{T}\mbh}W_{h}+\frac{1}{\lambda_{k}}(1-W_{h})\},
\end{eqnarray*}
where $W_{h}=\frac{\mbh^{T}\mbh}{\mbh^{T}\mbh+\mbh_{0}^{T}\mbho}$
is the weight between 0 and 1. Because $\log()$ is concave, we have
$\log(aW_{h}+b(1-W_{h}))\geq W_{h}\log(a)+(1-W_{h})\log(b)$. Hence,
\begin{eqnarray*}
f(\mbh,\mbho) & \geq & W_{h}\left\{ \log\frac{\mbh^{T}\bolPhi\mbh}{\mbh^{T}\mbh}+\log\frac{\mbh^{T}\bolOmega^{-1}\mbh}{\mbh^{T}\mbh}\right\} +(1-W_{h})\left\{ \log(\lambda_{k})+\log(\frac{1}{\lambda_{k}})\right\} \\
 & = & W_{h}\left\{ \log\frac{\mbh^{T}\bolPhi\mbh}{\mbh^{T}\mbh}+\log\frac{\mbh^{T}\bolOmega^{-1}\mbh}{\mbh^{T}\mbh}\right\} \\
 & \geq & W_{h}\cdot\min_{\mbh\in\mbbR^{d}}\left\{ \log\frac{\mbh^{T}\bolPhi\mbh}{\mbh^{T}\mbh}+\log\frac{\mbh^{T}\bolOmega^{-1}\mbh}{\mbh^{T}\mbh}\right\} \\
 & \geq & \min_{\mbh\in\mbbR^{d}}\left\{ \log\frac{\mbh^{T}\bolPhi\mbh}{\mbh^{T}\mbh}+\log\frac{\mbh^{T}\bolOmega^{-1}\mbh}{\mbh^{T}\mbh}\right\}.
\end{eqnarray*}
The last inequality holds because
\begin{equation}
\min_{\mbh\in\mbbR^{d}}\left\{ \log\frac{\mbh^{T}\bolPhi\mbh}{\mbh^{T}\mbh}+\log\frac{\mbh^{T}\bolOmega^{-1}\mbh}{\mbh^{T}\mbh}\right\} <0,\label{appendix min less zero}
\end{equation}
which is proved in Section~\ref{proof min less zero}.

Moreover, the lower bound of $f(\mbh,\mbho)$, which is negative,
will be attained if we let $W_{h}=1$ and let $\mbh=\arg\min_{\mbh\in\mbbR^{d}}\left\{ \log\frac{\mbh^{T}\bolPhi\mbh}{\mbh^{T}\mbh}+\log\frac{\mbh^{T}\bolOmega^{-1}\mbh}{\mbh^{T}\mbh}\right\} $.
So we have the minimum found at $W_{h}=\frac{\mbh^{T}\mbh}{\mbh^{T}\mbh+\mbh_{0}^{T}\mbho}=1$,
or equivalently, $\mbg=\bolGamma\mbh\in\spn(\bolGamma)$.

For the $(k+1)$-th direction, $\mbg_{k+1}=\mbG_{0k}\mbw_{k+1}$ where
$\mbw_{k+1}=\arg\min_{\mbw\in\mbbR^{p-k}}J_{k}(\mbw)$, $\mathrm{subject\ to}\ \mbw^{T}\mbw=1.$
Because $J_{k}(\mbw)=\log|\mbw^{T}\mbG_{0k}^{T}\mbM\mbG_{0k}\mbw|+\log|\mbw^{T}\left\{ \mbG_{0k}^{T}(\mbM+\mbU)\mbG_{0k}\right\} ^{-1}\mbw|$
has the same form as $f(\mbg)$, analogous to the first direction,
this gives $\mbw_{k+1}\in\calE_{\mbG_{0k}^{T}\mbM\mbG_{0k}}(\mbG_{0k}^{T}\calB)$.
Therefore $\mbg_{k+1}=\mbG_{0k}\mbw_{k+1}\in\calE_{\mbM}(\calB)$
by Proposition \ref{pn: seq breakdown envelope}.

\subsubsection{Proof of inequality (\ref{appendix min less zero})\label{proof min less zero}}
We first show that $\min_{\mbh\in\mbbR^{d}}\left\{ \log\frac{\mbh^{T}\bolPhi\mbh}{\mbh^{T}\mbh}+\log\frac{\mbh^{T}\bolOmega^{-1}\mbh}{\mbh^{T}\mbh}\right\} \leq0$, then we assume the equality to conduct the proof by contradiction.
Define the following two functions,
\begin{eqnarray*}
\F(\mbh;\bolPhi,\bolOmega^{-1})
&:=&\log\frac{\mbh^{T}\bolPhi\mbh}{\mbh^{T}\mbh}
+\log\frac{\mbh^{T}\bolOmega^{-1}\mbh}{\mbh^{T}\mbh},\\
\F(\mbh;\bolOmega,\bolOmega^{-1})
&:=&
\log\frac{\mbh^{T}\bolOmega\mbh}{\mbh^{T}\mbh}
+\log\frac{\mbh^{T}\bolOmega^{-1}\mbh}{\mbh^{T}\mbh},
\end{eqnarray*}
Recall that $\bolOmega-\bolPhi\geq0$, hence $\F(\mbh;\bolPhi,\bolOmega^{-1})\leq\F(\mbh;\bolOmega,\bolOmega^{-1})$ for any $\mbh$.
Consider the minimum of both $\F(\mbh;\bolPhi,\bolOmega^{-1})$ and $\F(\mbh;\bolOmega,\bolOmega^{-1})$, we have
\[
\min_{\mbh}\F(\mbh;\bolPhi,\bolOmega^{-1})
\leq
\min_{\mbh}\F(\mbh;\bolOmega,\bolOmega^{-1})=0,
\]
where the minimum of the right hand side is zero by taking $\mbh$
equals to any eigenvector of $\bolOmega$.

Now we assume that $\min_{\mbh}\F(\mbh;\bolPhi,\bolOmega^{-1})=0$.
Then for an arbitrary $\mbh$,
\begin{equation*}
0\leq \F(\mbh;\bolPhi,\bolOmega^{-1})
\leq
\F(\mbh;\bolOmega,\bolOmega^{-1}).
\end{equation*}
Let $\mbh_i=\mbell_i(\bolOmega)$, $i=1,\dots,u$, be the $i$-th unit eigenvector of $\bolOmega$ and plug $\mbh_i$ into the above inequalities, we have
\[
0\leq \F(\mbh_i;\bolPhi,\bolOmega^{-1})
\leq
\F(\mbh_i;\bolOmega,\bolOmega^{-1})=0,\ i=1,\dots,u,
\]
which implies
\[
0= \F(\mbh_i;\bolPhi,\bolOmega^{-1})
=
\F(\mbh_i;\bolOmega,\bolOmega^{-1})=0,\ i=1,\dots,u,
\]
and more explicitly,
\[
\log(\mbh_i^T\bolPhi\mbh_i)=\log(\mbh_i^T\bolOmega\mbh_i),\ i=1,\dots,u,
\]
which implies $\bolPhi=\bolOmega$ because that $\bolPhi,\ \bolOmega\in\mbbR^{u\times u}$ and $\mbh_i$, $i=1,\dots,u$, are $u$ linear independent vectors.
Then by definition $\mbU=\bolGamma^T(\bolPhi-\bolOmega)\bolGamma=0$ leads to contradiction with the dimension of the envelope.

\end{proof}

\subsection{Proposition \ref{pn: sqrt_n_consistent 1D manifold algorithm}}\label{supp subsec: root-n 1D}

\begin{proof}

Our proof of $\sqrt{n}$-consistency hinges on Amemiya's
(1985) results on the asymptotic properties of extremum estimators.
Proposition 4.1.1 and Proposition 4.1.3 in Amemiya (1985) can be applied to
our context. We first state these results and then sketch how they
can be used to prove the $\sqrt{n}$-consistency for our algorithm.

Let $Q_{n}(\mby,\boltheta)$ be a real-valued function of the random
variables $\mby=(\mby_{1},\dots,\mby_{n})^{T}$ and the parameters
$\boltheta=(\boltheta_{1},\dots,\boltheta_{K})^{T}$. We shall sometimes
write $Q_{n}(\mby,\boltheta)$ more compactly as $Q_{n}(\boltheta)$.
Let the parameter space be $\bolTheta$ and let the true value of
$\boltheta$ be $\boltheta_{t}$ which is in $\bolTheta$. Then Proposition
4.1.1 and Proposition 4.1.3 in Amemiya (1985) give asymptotic properties of the extremum estimator,
$\widehat{\boltheta}_{n}=\arg\max_{\boltheta\in\bolTheta}Q_{n}(\mby,\boltheta)$.
We summarize the conditions in Amemiya's Propositions as follows.

\begin{itemize}
\item[(A)]The parameter space $\bolTheta$ is a compact subset of $\mbbR^{K}$;
\item[(B)]$Q_{n}(\mby,\boltheta)$ is continuous in $\boltheta\in\bolTheta$;
for all $\mby$ and is a measurable function of $\mby$ for all $\boltheta\in\bolTheta$;
\item[(C)]$n^{-1}Q_{n}(\boltheta)$ converges to a nonstochastic function
$Q(\boltheta)$ in probability uniformly in $\boltheta\in\bolTheta$
as $n$ goes to infinity, and $Q(\boltheta)$ attains a unique global
maximum at $\boltheta_{t}$;
\item[(D)]$\partial^{2}Q_{n}(\boltheta)/\partial\boltheta\partial\boltheta^{T}$
exists and is continuous in an open, convex neighborhood of $\boltheta_{0}$;
\item[(E)]$n^{-1}\left\{ \partial^{2}Q_{n}(\mby,\boltheta)/\partial\boltheta\partial\boltheta^{T}\right\} _{\boltheta=\boltheta_{n}^{*}}$
converges to a finite nonsingular matrix
\[
\mbA(\boltheta_{t})=\lim_{n\rightarrow\infty}\E_{\boltheta_{t}}\left\{ n^{-1}\left\{ \partial^{2}Q_{n}(\boltheta)/\partial\boltheta\partial\boltheta^{T}\right\} \right\} ,
\]
for any random sequences $\boltheta_{n}^{*}$ such that $\mathrm{plim}(\boltheta_{n}^{*})=\boltheta_{t}$;
\item[(F)]$n^{-1/2}\left\{ \partial Q_{n}(\boltheta)/\partial\boltheta\right\} _{\boltheta=\boltheta_{t}}\rightarrow N(0,\mbB(\boltheta_{t}))$,
where
\[
\mbB(\boltheta_{t})=\lim_{n\rightarrow\infty}\E_{\boltheta_{t}}\left\{ n^{-1}\left\{ \partial Q_{n}(\boltheta)/\partial\boltheta\right\} \left\{ \partial Q_{n}(\boltheta)/\partial\boltheta^{T}\right\} \right\}.
\]
\end{itemize}

\begin{pn}\label{pn: Amemiya 4.1.1} Under assumptions (A)-(C),
$\widehat{\boltheta}_{n}$ converges to $\boltheta_{t}$ in probability.
\end{pn}
\begin{pn}\label{pn: Amemiya 4.1.3} Under assumptions (A)-(F),
$\sqrt{n}(\widehat{\boltheta}_{n}-\boltheta_{t})\rightarrow N(0,\mbA(\boltheta_{t})^{-1}\mbB(\boltheta_{t})\mbA(\boltheta_{t})^{-1}).$
\end{pn}

In our adaptation of Proposition \ref{pn: Amemiya 4.1.1} and Proposition
\ref{pn: Amemiya 4.1.3}, we let $\boltheta\equiv\mbg$ whose true
value is denoted by $\mbg_{t}$ and let the random variables $\mby=\vech(\widehat{\mbM},\widehat{\mbU})$.
The parameter space is the 1D manifold $\bolTheta=\mathcal{G}_{(p,1)}$
which is a compact subset of $\mbbR^{p}$, so condition $(A)$
in Proposition \ref{pn: Amemiya 4.1.1} is satisfied. The function to
be maximized is defined as follows.
\begin{equation}
Q_{n}(\mbg)=-n/2\log(\mbg^{T}\widehat{\mbM}\mbg)-n/2\log(\mbg^{T}(\widehat{\mbM}+\widehat{\mbU})^{-1}\mbg)+n\log(\mbg^{T}\mbg).\label{Qn theta}
\end{equation}
Condition $(B)$ then holds. We next verify condition $(C)$ that
$n^{-1}Q_{n}(\mbg)$ converges uniformly to
\begin{equation}
Q(\mbg)=-1/2\log(\mbg^{T}\mbM\mbg)-1/2\log(\mbg^{T}(\mbM+\mbU)^{-1}\mbg)+\log(\mbg^{T}\mbg).\label{Q theta}
\end{equation}
We have shown that the population objective function $Q(\mbg)$ attains
the unique global maximum at $\mbg_{t}$. For simplicity, we assume
$\mbM$ and $\mbM+\mbU$ both have distinct eigenvalues so that $\mbg_{t}$
is the unique maximum of $Q(\mbg)$ in the 1D manifold $\bolTheta$.
For the case where there are multiple local maxima of $Q(\mbg)$,
we can obtain similar results by applying Proposition 4.1.2 in Amemiya
(1985) as an alternative of Proposition \ref{pn: Amemiya 4.1.1}. Since
$\widehat{\mbM}$ and $\widehat{\mbU}$ are $\sqrt{n}$-consistent
for $\mbM$ and $\mbU$, the eigenvectors and eigenvalues of
$\widehat{\mbM}$ and $(\widehat{\mbM}+\widehat{\mbU})^{-1}$ are $\sqrt{n}$-consistent
for the eigenvectors and eigenvalues of their population counterparts.

Then $n^{-1}Q_{n}(\mbg)$ converge in probability to $Q(\mbg)$ uniformly
in $\mbg$, as can be seen from the following argument.
\begin{eqnarray*}
n^{-1}Q_{n}(\mbg)-Q(\mbg) & = & -1/2\left(\log(\mbg^{T}(\widehat{\mbM}+\widehat{\mbU})^{-1}\mbg)-\log(\mbg^{T}(\mbM+\mbU)^{-1}\mbg)\right)\\
 &  & -1/2\left(\log(\mbg^{T}\widehat{\mbM}\mbg)-\log(\mbg^{T}\mbM\mbg)\right)\\
 & = & -1/2\log\left[\frac{\mbg^{T}(\widehat{\mbM}+\widehat{\mbU})^{-1}\mbg}{\mbg^{T}(\mbM+\mbU)^{-1}
 \mbg}\right]-1/2\log\left[\frac{\mbg^{T}\widehat{\mbM}\mbg}{\mbg^{T}\mbM\mbg}\right].
\end{eqnarray*}
Hence, $\sup_{\mbg\in\bolTheta}\log(\mbg^{T}\widehat{\mbM}\mbg/\mbg^{T}\mbM\mbg)
=\sup_{\mbg\in\bolTheta}\log(\mbg^{T}\mbM^{-1/2}\widehat{\mbM}\mbM^{-1/2}\mbg/\mbg^{T}\mbg)$,
which equals to the logarithm of the largest eigenvalue of $\mbM^{-1/2}\widehat{\mbM}\mbM^{-1/2}$
and converges to 0 in probability. Similarly, $\sup_{\mbg\in\bolTheta}\log[\mbg^{T}(\widehat{\mbM}+\widehat{\mbU})^{-1}\mbg/\mbg^{T}(\mbM+\mbU)^{-1}\mbg]$
converges to zero in probability. Therefore, $n^{-1}Q_{n}(\mbg)$
converges to $Q(\mbg)$ in probability uniformly in $\mbg\in\bolTheta$.
Note that we have assumed $\mbM+\mbU>0$ and $\mbM^{-1}>0$, so their
eigenvalues will be bounded away from zero.

We next verify conditions $(D)-(F)$. By straightforward calculation,
condition $(D)$ follows from the second derivative matrix
\begin{eqnarray}
n^{-1}\frac{\partial^{2}Q_{n}(\mbg)}{\partial\mbg\partial\mbg^{T}} & = & 2(\mbg^{T}\widehat{\mbM}\mbg)^{-2}(\widehat{\mbM}\mbg\mbg^{T}\widehat{\mbM})
-(\mbg^{T}\widehat{\mbM}\mbg)^{-1}\widehat{\mbM}\nonumber \\
 &  & +2\left[\mbg^{T}(\widehat{\mbM}+\widehat{\mbU})^{-1}\mbg\right]^{-2}
 \left[(\widehat{\mbM}+\widehat{\mbU})^{-1}\mbg\mbg^{T}(\widehat{\mbM}+\widehat{\mbU})^{-1}
 \right]\nonumber \\
 &  & -\left[\mbg^{T}(\widehat{\mbM}+\widehat{\mbU})^{-1}\mbg\right]^{-1}(\widehat{\mbM}+
 \widehat{\mbU})^{-1}\nonumber\\
 &&-2(\mbg^{T}\mbg)^{-2}\mbP_{\mbg}+(\mbg^{T}\mbg)^{-1}\mbI_{p}.\label{verify conditionE in Amemiya}
\end{eqnarray}
Condition $(E)$ holds because the above quantity is a smooth function
of $\mbg$, $\widehat{\mbM}$ and $(\widehat{\mbM}+\widehat{\mbU})^{-1}$.

Last, we need to verify condition $(F)$. From the proof of Proposition
\ref{pn: Amemiya 4.1.3}, we need only show that $n^{-1}\left\{ \partial Q_{n}(\boltheta)/\partial\boltheta\right\} _{\boltheta=\boltheta_{0}}=O_{p}(1/\sqrt{n})$
for $\sqrt{n}$-consistency of the estimator $\widehat{\boltheta}_{n}$.
The derivative $n^{-1}\left\{ \partial Q_{n}(\mbg)/\partial\mbg\right\} _{\mbg=\mbg_{t}}$
equals
\begin{equation}
-(\mbg_{t}^{T}\widehat{\mbM}\mbg_{t})^{-1}\widehat{\mbM}\mbg_{t}
-(\mbg_{t}^{T}(\widehat{\mbM}+\widehat{\mbU})^{-1}\mbg_{t})^{-1}(\widehat{\mbM}+\widehat{\mbU})^{-1}\mbg_{t}+2\mbg_{t}.
\end{equation}
Following the derivation for the population objective function, we know
that $\left\{ \partial Q(\mbg)/\partial\mbg\right\} _{\mbg=\mbg_{t}}=0$.
Then the result follows from the fact that $n^{-1}\partial Q_{n}(\mbg)/\partial\mbg$
is a smooth function of $\widehat{\mbM}$ and $(\widehat{\mbM}+\widehat{\mbU})^{-1}$
which are $\sqrt{n}$-consistent estimators.

So far, we have verified the conditions $(A)-(F)$ so that the sample
estimator $\widehat{\mbg}_{1}$ will be $\sqrt{n}$-consistent for
the population estimator. For the ($k+1$)-th direction, $k<u$, let
$\widehat{\mbG}_{k}$ denote an $\sqrt{n}$-consistent estimator of
the first $k$ directions and let $(\widehat{\mbG}_{k},\widehat{\mbG}_{0k})$
be an orthogonal matrix. The ($k+1$)-th direction is defined by $\mbg_{k+1}=\widehat{\mbG}_{0k}\mbw_{k+1}$
where the parameters are $\mbw_{k+1}\in\bolTheta_{k+1}\subset\mbbR^{p-k}$
and the parameter space is $\bolTheta_{k+1}=\mathcal{G}_{p-k,1}$.
We show that we can obtain a $\sqrt{n}$-consistent estimator $\widehat{\mbw}_{k+1}$,
so the $\sqrt{n}$-consistency of $\widehat{\mbg}_{k+1}=\widehat{\mbG}_{0k}\widehat{\mbw}_{k+1}$
then follows. We define our objective functions $Q_{n}(\mbw)$ and
$Q(\mbw)$ as
\begin{align*}
Q_{n}(\mbw)  & =  -n/2\log(\mbw^{T}(\widehat{\mbG}_{0k}^{T}(\widehat{\mbM}+\widehat{\mbU})\widehat{\mbG}_{0k})^{-1}\mbw)-n/2\log(\mbw^{T}\widehat{\mbG}_{0k}^{T}\widehat{\mbM}\widehat{\mbG}_{0k}\mbw)+n\log(\mbw^{T}\mbw)\\
Q(\mbw) & =  -1/2\log(\mbw^{T}(\mbG_{0k}^{T}(\mbM+\mbU)\mbG_{0k})^{-1}\mbw)-1/2\log(\mbw^{T}\mbG_{0k}^{T}\mbM\mbG_{0k}\mbw)+\log(\mbw^{T}\mbw)
\end{align*}
Following the same logic as verifying the conditions for the first
direction, we can see that $\widehat{\mbw}=\arg\max Q_{n}(\mbw)$
will be $\sqrt{n}$-consistent for $\mbv_{t}=\arg\max Q(\mbw)$ by
noticing that $(\widehat{\mbG}_{0k}^{T}(\widehat{\mbM}+\widehat{\mbU})\widehat{\mbG}_{0k})^{-1}$
and $\widehat{\mbG}_{0k}^{T}\widehat{\mbM}\widehat{\mbG}_{0k}$
are $\sqrt{n}$-consistent estimators for $(\mbG_{0k}^{T}(\mbM+\mbU)\mbG_{0k})^{-1}$
and $\mbG_{0k}^{T}\mbM\mbG_{0k}$. Since all the $u$ directions
will be $\sqrt{n}$-consistent, the projection onto $\widehat{\mbG}_{u}=(\widehat{\mbg}_{1},\dots,\widehat{\mbg}_{u})$
will be a $\sqrt{n}$-consistent estimator for the projection onto
the envelope $\calE_{\mbM}(\mbU)$.
\end{proof}

\end{appendix}


\begin{thebibliography}{References}

\bibitem{key-1} \textsc{P. A. Absil, R. Mahony, and R. Sepulchrer} (2008), Optimization Algorithms on Matrix Manifolds. {\em Princeton University Press}.

\bibitem{key-1011} \textsc{Amemiya, T.} (1985), Advanced Econometrics, {\em Harvard University Press}.



%
%


\bibitem{key-3} \textsc{Conway, J.} (1990). A Course in Functional Analysis. Second edition. {\em Springer, New York}.
%
%

\bibitem{key-3a} \textsc{Cook, R.D., Helland, I.S. and Su, Z.} (2013), Envelopes and partial least squares regression. {\em JRSS-B}, \textbf{75},851--877.

\bibitem{key-4}\textsc{ Cook, R.D., Li, B. and Chiaromonte, F. } (2010). Envelope models for parsimonious and efficient multivariate linear
regression (with discussion). {\em Statistica Sinica}, \textbf{20},927--1010.

\bibitem{key-11f} \textsc{Cook, R.D. and Zhang, X.} (2014). Simultaneous envelopes for multivariate linear regression. {\em Technometrics.} DOI:10.1080/00401706.2013.872700

\bibitem{key-11b} \textsc{Lehmann, E. L. and Casella, G.} (1998). Theory of Point Estimation. Second edition. {\em Springer, New York}.

\bibitem{key-11c} \textsc{de Jong, S.} (1993), SIMPLS: an alternative approach to partial least squares regression. {\em Chemometr. Intell. Lab. Syst.},\textbf{18}, 251--26.

\bibitem{key-11d} \textsc{Kenward, M. G.} (1987),  A method for comparing profiles of repeated measurements. {\em JRSS-C}, \textbf{36}, 296--308.

%





%

\bibitem{key-19} \textsc{Seber, G.A.F.} (2008), A matrix handbook for statisticians, {\em Wiley-Interscience}.


\bibitem{key-15} \textsc{Su, Z. and Cook, R.D. } (2011), Partial envelopes
for efficient estimation in multivariate linear regression. {\em
Biometrika}, \textbf{98}, 133--146.



\end{thebibliography}
\end{document}